\documentclass[aps,prb,twocolumn,amsmath,amssymb,superscriptaddress,floatfix]{revtex4}
\usepackage{graphicx}
\usepackage{graphics}
\usepackage{bm}
\usepackage[usenames]{color}
\usepackage{colordvi}
\usepackage{units}
\usepackage{bbm}
\usepackage{booktabs}
\usepackage{array}
\usepackage{placeins}
\usepackage{epsfig}
\usepackage{lscape}
\usepackage{changes}
\usepackage{braket}
\usepackage{tabularx}

\newcolumntype{L}[1]{>{\raggedright\arraybackslash}p{#1}} 
\newcolumntype{C}[1]{>{\centering\arraybackslash}p{#1}} 
\newcolumntype{R}[1]{>{\raggedleft\arraybackslash}p{#1}} 

\bibstyle{apsrev.bib}

\begin{document}

\newcommand{\be}{\begin{equation}}
\newcommand{\ee}{\end{equation}}
\newcommand{\beqn}{\begin{eqnarray}}
\newcommand{\eeqn}{\end{eqnarray}}

\title{Quantum XX-model with competing short- and long-range interactions:\\ Phases and phase transitions in and out of equilibrium}  
\author{Ferenc Igl{\'o}i}
\email{igloi.ferenc@wigner.mta.hu}
\affiliation{Wigner Research Centre for Physics, Institute for Solid State Physics and Optics, H-1525 Budapest, P.O. Box 49, Hungary}
\affiliation{Institute of Theoretical Physics, Szeged University, H-6720 Szeged, Hungary}
\affiliation{Theoretische Physik, Saarland University, D-66123 Saarbr{\"u}cken, Germany}
\author{Benjamin Bla{\ss}}
\email{bebla@lusi.uni-sb.de}
\affiliation{Theoretische Physik, Saarland University, D-66123 Saarbr{\"u}cken, Germany}
\author{Gerg\H o Ro\'osz}
\email{roosz.gergo@wigner.mta.hu}
\affiliation{Wigner Research Centre for Physics, Institute for Solid State Physics and Optics, H-1525 Budapest, P.O. Box 49, Hungary}
\affiliation{Institute of Theoretical Physics, Szeged University, H-6720 Szeged, Hungary}
\affiliation{Institute of Theoretical Physics, Technische Universit\"at Dresden, D-01062 Dresden, Germany}
\author{Heiko Rieger}
\email{h.rieger@physik.uni-saarland.de}
\affiliation{Theoretische Physik, Saarland University, D-66123 Saarbr{\"u}cken, Germany}

\date{\today}

\begin{abstract}
We consider the quantum XX-model in the presence of competing nearest-neighbour and global-range interactions, which is equivalent to a Bose-Hubbard model with cavity mediated global range interactions in the hard core boson limit. Using fermionic techniques the problem is solved exactly in one dimension in the thermodynamic limit. The ground state phase diagram consists of two ordered phases: ferromagnetic (F) and antiferromagnetic (AF), as well as an XY-phase having quasi-long-range order. We have also studied quantum relaxation after sudden quenches. Quenching from the AF phase to the XY region remanent AF order is observed below a dynamical transition line. In the opposite quench, from the XY region to the AF-phase beyond a static metastability line AF order arises on top of remanent XY quasi-long-range order, which corresponds to dynamically generated supersolid state in the equivalent Bose-Hubbard model with hard-core bosons.

\end{abstract}

\pacs{}

\maketitle

\section{Introduction}
\label{sec:intr}

Recently there is an increased interest to study the phase diagram and non-equilibrium dynamics of quantum many-body systems with competing short- and long-range interactions. Experimentally, such systems have been realised with ultracold atoms in optical lattices inside a high-finesse optical cavity\cite{Science-Esslinger,Klinder2015,Landig2016}. The strength of the short-range (on-site) interaction is related to the depth of the optical lattice, while long- (infinite) range interactions are controlled by a vacuum mode of the cavity. The interplay of short- and long-range interactions may result in a rich phase-diagram with exotic phases and interesting non-equilibrium dynamics. 

Classical many-body systems with competing short- and ferromagnetic global-range interactions have been studied earlier and in the thermodynamic limit the order-parameter is obtained through a self-consistent treatment, like in mean-field models. \cite{Kardar1983,Mukamel2005}. Theoretical results for the
phase diagrams of 
quantum many body systems with competing short-range and global-range interactions are rare and up to now confined to the aforementioned 
bosonic system\cite{Chen2016,Dogra2016,Niederle2016,Sundar2016,Panas2017,Flottat2017}.

The non-equilibrium dynamics of closed quantum many body systems 
after sudden quenches has attracted a lot of attention in the last 
decade. Here one is interested in the time-evolution of different observables, such as the order parameter or some correlation function, after the quench. Fundamental questions concerning quantum quenches include i) the functional form of the relaxation process in early times, and ii) the properties of the stationary state of the system after sufficiently long time. The latter problem is related to the question of thermalization, which is expected to be different for integrable and non-integrable quantum systems.
Non-integrable systems are expected to evolve into a thermalized state, 
in which local observables can be characterized by thermal expectation values~\cite{Rigol_07,Calabrese_06,Calabrese_07,Cazalilla_06,Manmana_07,
Cramer_08,Barthel_08,Kollar_08,Sotiriadis_09,Roux_09,Sotiriadis_11}, 
but there are some counterexamples~\cite{larson,hamazaki,blass_rieger}. On the other hand,
integrable systems in the stationary state are generally described by a
so-called generalized Gibbs ensemble, which includes all the conserved
quantities of the system. Recently it has been observed that in certain models
both local and quasi-local conserved quantities
have to be
taken into account to construct an appropriate
generalized Gibbs ensemble~\cite{wouters,pozsgay,goldstein,pozsgay1,pozsgay2,essler_mussardo_panfil,ilievski1,ilievski2,doyon,vidmar}.

Sudden quenches have been studied for bosons in optical lattices 
experimentally \cite{BH-dynamics-exp} and theoretically 
\cite{BH-dymamics-theory}. The underlying Hamiltonian, the
Bose-Hubbard model with nearest-neighbour interactions, is known to be non-integrable, and thus the 
dynamics expected to thermalize, but numerical studies,
comprising DMRG in one dimension \cite{DMRG}, t-VMC in 
higher dimensions \cite{tVMC} or
numerical dynamical MFT \cite{dMFT}
indicate non-thermal behaviour for strong quenches. Non-equilibruim dynamics of the bosonic system in the cavity-setup
has been studied recently and after very long times the dynamics is observed to become incoherent, which is explained that dissipation due to photon loss can take place 
\cite{Esslinger-Metastab,QMC-trap}.

In this paper we study theoretically quantum many-body systems with competing short- and long- (infinite) range interactions. The system we consider is the quantum XX-model, which is closely related to the Bose-Hubbard model. 
For hard core bosons, when the lattice sites have only single occupation the boson operators can be represented by spin-$1/2$ operators. In the actual calculation the quantum XX-model is put on a one dimensional lattice and we solve its ground state and quench dynamics exactly by free-fermionic techniques.
A brief account of our results translated into the bosonic language has been 
given by us recently \cite{Blass-PRL}. 

The paper is organised as follows. In Sec.\ref{sec:model} we present the model and show, how the term with global-range interaction is transformed to an effective field, the strength of which is calculated self-consistently in the thermodynamic limit. In Sec.\ref{sec:free_fermion} we solve the one-dimensional model exactly and in Sec.\ref{Sec:phase_diagram} we calculate the quantum phases and phase transitions in the ground state. In Sec.\ref{sec:dynamics} we study non-equilibrium dynamics of the model after a quench and properties of dynamical phase transitions are calculated. Our results are discussed in Sec.\ref{sec:discussion} and detailed derivation of different results are put in the Appendices.

\section{The quantum XX-model with global range interactions}
\label{sec:model}

Let us consider the quantum XX-model with global range interactions in a one-dimensional lattice, defined by the Hamiltonian:
\begin{align}
\begin{split}
\hat{H}=&-J\sum_{j=1}^L\left(\sigma_{j}^{x} \sigma_{j+1}^{x}+\sigma_{j}^{y}\sigma_{j+1}^{y}\right)\\
&-h\sum_{j=1}^L \sigma^z_{j}
-\varepsilon\frac{1}{L}\left(\sum_{j,odd}\sigma^z_{j}-\sum_{j,even}\sigma^z_{j}\right)^2\;.
\label{XX-Ham1}
\end{split}
\end{align}
Here the $\sigma_{j}^{x,y,z}$ are Pauli matrices at site $i$, the nearest neighbour coupling constant and the strength of the transverse field are denoted by $J$ and $h$, respectively. The last term of the r.h.s. represents 
a global range antiferromagnetic interaction of strength $\varepsilon$,
favoring anti-parallel $z$-orientation of the spins on even and odd lattice sites.  
It is expressed as the square of the staggered magnetization operator:
\begin{align}
\hat{x}=\frac{1}{L}\left(\sum_{j,odd}\sigma^z_{j}-\sum_{j,even}\sigma^z_{j}\right)\;.
\label{stag_mg}
\end{align}
We note that the Hamiltonian in Eq.(\ref{XX-Ham1}) is equivalent to the Bose-Hubbard model for hard core bosons, which is explained in Appendix A. Here the global range interaction represents a cavity mediated long
range interaction \cite{Cavity_review,BH-Cavity1,BH-Cavity2}
and therefore has immediate experimental relevance\cite{Science-Esslinger,Klinder2015,Landig2016}.
Consequently our results on the XX-model can be translated to 
hard core lattice bosons in 1d with cavity mediated long range
interactions \cite{Blass-PRL}.

In the next step we transform the Hamiltonian in Eq.(\ref{XX-Ham1}) in an equivalent form, by linearizing the global-range interaction term in the thermodynamic limit. Following the steps of the derivation in the Appendix we arrive to the Hamiltonian:
\begin{align}
\begin{split}
\hat{H'}(x)=&-J\sum_{j=1}^L\left(\sigma_{j}^{x} \sigma_{j+1}^{x}+\sigma_{j}^{y}\sigma_{j+1}^{y}\right)
-h\sum_{j=1}^L \sigma^z_{j}\\
&-2x \varepsilon \left(\sum_{j,odd}\sigma^z_{j}-\sum_{j,even}\sigma^z_{j}\right)+L\varepsilon x^2\;,
\label{XX-Ham2}
\end{split}
\end{align}
where $x$ has to be determined self-consistently via
\begin{equation}
x=\langle\hat x\rangle
=\frac1L\biggl\langle\sum_{j,odd}\sigma_i^z -\sum_{j,even}\sigma_i^z\biggr\rangle_{H'(x)}\;.
\label{SK}
\end{equation}
$\langle\ldots\rangle_{H'}$ is the average in the ground state of the system defined by $H'$. This condition is equivalent to the requirement that the ground-state energy of $\hat{H'}(x)$ is minimal with respect of $x$, which follows from the Hellmann-Feynmann theorem:
\begin{equation}
\frac{{\rm d} E_{0}(x)}{{\rm d} x}= \left<\frac{{\rm d} {H'}(x)}{{\rm d} x} \right>=-2L\epsilon[\left<{\hat x}\right>-x]=0\;.
\label{minimum}
\end{equation}
At this point we make two comments about possible generalisation of the treatment of the cavity-induced global-range interaction term. First, the equivalence of the two Hamiltonians $\hat{H}$ and $\hat{H}'(x)$ with (\ref{SK}) is valid for bipartite lattices in arbitrary dimensions in the thermodynamic limit $N\to\infty$. Second, using the technique of Appendix B a cavity induced global-range interaction term can be linearised in the thermodynamic limit for other models, too, such as for the extended Bose-Hubbard model (with soft- or hard-core bosons), as defined in Eq.(\ref{BH-Ham}) the Appendix A, and also
to fermionic models as for instance free fermions on bipartite lattices. 

\section{Free-fermion solution}
\label{sec:free_fermion}

Using the Jordan-Wigner transformation the Hamiltonian in Eq.(\ref{XX-Ham2}) is transformed in terms
of the fermion creation ($c^{\dag}_j$) and annihilation ($c_j$) operators in quadratic form:
\begin{align}
\begin{split}
{\cal H'}&=-2J\sum_{j=1}^{L-1} \left( c^{\dag}_j c_{j+1} + c^{\dag}_{j+1} c_{j} \right)+2J w \left(c^{\dag}_L c_{1} +c^{\dag}_{1} c_{L} \right)\\
&- \sum_{j=1}^L 2 (h+ 2\epsilon x e^{i\pi j} )(c^{\dag}_j c_{j}-1/2)+L\epsilon x^2\;,
\label{hamiltonian_mf1}
\end{split}
\end{align}
with $w=\exp(i\pi N_c)$ and $N_c=\sum_{j=1}^L c^{\dag}_j c_{j}$.
In the next step we introduce the Fourier representation:
\begin{equation}
c_j=\frac{1}{\sqrt{L}}\sum_k c_k e^{-\imath k j};,
\label{fourier}
\end{equation}
where the $k$-values are in the range: $-\pi < k < \pi$. 
(For $w=1$ these are $k=\pm \frac{(2j-1)\pi}{L}$, $j=1,2,\dots L/2$, while for $w=-1$ these are $k=0,\pm \frac{2j\pi}{L},\pi$, with $j=1,2,\dots L/2-1$)
In terms of the Fourier operators the Hamiltonian assumes the form:
\begin{align}
\begin{split}
{\cal H'} &= \sum_{k>0}{\cal H}_{k}\;,\\
{\cal H}_{k}&= -2(h+2J\cos k) c^{\dag}_k c_k-2(h-2J\cos k)c^{\dag}_{k-\pi} c_{k-\pi}\\
 &+ 4\epsilon x(c^{\dag}_{k} c_{k-\pi}+c^{\dag}_{k-\pi} c_{k})+2h+2\epsilon x^2\;,
\label{hamiltonian_mf2}
\end{split}
\end{align}
which is separated into $L/2$ independent $2 \times 2$ sectors.

The ${\cal H}_k$ operators are diagonalized by the canonical transformation:
\begin{align}
\begin{split}
\eta_k&=g_{k,k}c_k+g_{k,k-\pi}c_{k-\pi}\\
\eta_{k-\pi}&=g_{k-\pi,k}c_k+g_{k-\pi,k-\pi}c_{k-\pi} \;,
\label{eta}
\end{split}
\end{align}
with
\begin{align}
\begin{split}
g_{k,k}&=g_{k-\pi,k-\pi}=\left[ 1 + \left(\sqrt{a_k^2+1}-a_k \right)^2 \right]^{-1/2}\\ 
g_{k,k-\pi}&=-g_{k-\pi,k}=-\left[ 1 + \left(\sqrt{a_k^2+1}+a_k \right)^2 \right]^{-1/2} \;,
\label{g}
\end{split}
\end{align}
and $a_k=\frac{J }{\epsilon x}\cos k$. Then:
\begin{equation}
{\cal H}_{k}= \Lambda_k (\eta^{\dag}_k \eta_k-1/2)+\Lambda_{k-\pi} (\eta^{\dag}_{k-\pi} \eta_{k-\pi}-1/2)+2\epsilon x^2\;,
\label{hamiltonian_mf3}
\end{equation}
with:
\begin{align}
\begin{split}
\Lambda_k&=-2h - 4\sqrt{J^2\cos^2 k + \epsilon^2 x^2}\\
\Lambda_{k-\pi}&=-2h + 4\sqrt{J^2\cos^2 k + \epsilon^2 x^2}\;.
\label{Lambda}
\end{split}
\end{align}
Note that the energy of modes is symmetric to $k=\pi/2$, thus we can restrict ourselves to the range: $0<k<\pi/2$, however with ${\cal H}_{k} \to 2{\cal H}_{k}$.
The inverse of Eq.(\ref{eta}) is given by:
\begin{align}
\begin{split}
c_k&=g_{k,k}\eta_k+g_{k-\pi,k}\eta_{k-\pi}\\ 
c_{k-\pi}&=g_{k,k-\pi}\eta_k+g_{k-\pi,k-\pi}\eta_{k-\pi} \;,
\label{eta1}
\end{split}
\end{align}
\section{Ground-state phase diagram}
\label{Sec:phase_diagram}

The energy of modes of the diagonalised Hamiltonian are $\Lambda_k < 0$ for all $k \in (0,\pi/2)$, but the
$\Lambda_{k-\pi}$ are positive, if $\cos^2 k > (\frac{h}{2J})^2-(\frac{\epsilon x}{J})^2$. In the following we characterise a state by a wavenumber $k_m$, so that $\braket{\hat{\eta}_{k}^{\dagger}\hat{\eta}_{k}}_{{k_m}}=1$
for all $k$ and $\braket{\hat{\eta}_{k-\pi}^{\dagger}\hat{\eta}_{k-\pi}}_{{k_m}}=0$
for $k\in(0,k_{m})$ and $1$ for $k\in(k_{m},\pi/2)$. The energy per site is given by: 
\begin{align}
\begin{split}
&e(k_m)=\frac{1}{L}\sum_{k\in\left(0,k_{m}\right)}\Lambda_{k}+
\frac{1}{L}\sum_{k\in\left(k_{m},\frac{\pi}{2}\right)}
(\Lambda_{k}+\Lambda_{k-\pi}) +\varepsilon x^{2}\\
&=-h\left(1-\frac{2k_m}{\pi}\right)-\frac{4}{\pi}\int_0^{k_m}{\rm d}k \sqrt{J^2\cos^2 k + \epsilon^2 x^2}+\epsilon x^2\;.
\label{e-km}
\end{split}
\end{align}
Here the second term of the r.h.s. of the last equation can be expressed as $-\frac{4}{\pi}\sqrt{J^2+\epsilon^2x^2}E(k_m,[\epsilon^2 x^2/J^2+1]^{-1/2})$, in terms of the elliptic integral of the second kind: $E(k_m,q)$.

The self-consistency criterion is obtained from Eq.(\ref{e-km}) through Eq.(\ref{minimum}):
\begin{align}
\begin{split}
x&=\frac{2\epsilon x}{\pi}\int_0^{k_m} {\rm d} k \frac{1}{\sqrt{J^2 \cos^2 k+\epsilon^2 x^2}}\\
&=\frac{2\epsilon x}{\pi\sqrt{J^2 + \epsilon^2 x^2}}F(k_m,[\epsilon^2 x^2/J^2+1]^{-1/2})\;,
\label{Self_c}
\end{split}
\end{align}
where $F(\phi,q)$ is the elliptic integral of the first kind. In the ground state $e_0=\min_{k_m} e(k_m)$.
We note that using the representation of the staggered magnetization:
\begin{align}
\hat{x}=\frac{1}{L}\sum_{k>0}\left(\hat{c}_{k}^{\dagger}\hat{c}_{k-\pi}
+\hat{c}_{k-\pi}^{\dagger}\hat{c}_{k}\right)\;.
\label{Eq:Imbalance_c}
\end{align}
the same self-consistency criterion can be obtained through Eq.(\ref{SK}).

The stability of the self-consistent solution depends on the sign of the second-derivative:
\begin{align}
\begin{split}
\frac{{\rm d}^2 e}{{\rm d}x^2}&=-\frac{4\epsilon^2}{\pi}\int_0^{k_m} {\rm d} k \frac{1}{\sqrt{J^2\cos^2 k+\epsilon^2 x^2}}\\
&+\frac{4\epsilon^4x^2}{\pi}\int_0^{k_m} {\rm d} k \frac{1}{(J^2\cos^2 k+\epsilon^2 x^2)^{3/2}}+2\epsilon\;.
\label{der2}
\end{split}
\end{align}
The trivial solution, $\tilde{x}=0$ represents a (local) minimum, if
\begin{align}
\frac{J}{\epsilon} > \frac{2}{\pi}\int_0^{k_m} {\rm d} k \frac{1}{\cos k}=\frac{2}{\pi}\ln\left[ \tan\left(\frac{\pi}{4}+\frac{k_m}{2}\right)\right]\;,
\end{align}
which is satisfied for $0 < k_m< \tilde{k}_m$, with
\begin{align}
\tilde{k}_m(\epsilon)=2 \arctan\left[\exp\left(\frac{\pi J}{2\epsilon}\right)\right]-\frac{\pi}{2}\;.
\end{align}
One can show similarly, that for $\tilde{k}_m<k_m \le \pi/2$ the non-trivial self-consistent solution, $\tilde{x}>0$, is also a (local) minimum. This follows from the fact, that for $\tilde{x}>0$ the first and third terms at the r.h.s. of Eq.(\ref{der2}) cancel and the remaining second term is positive.
We conclude that at a fixed value of $k_m$ there is always one stable self-consistent solution, which is the trivial one, $\tilde{x}=0$, in the first regime, $0 < k_m < \tilde{k}_m$, and it is the non-trivial one, $\tilde{x}>0$, in the second regime, $\tilde{k}_m<k_m \le \pi/2$.

For fixed values of the parameters, $J$, $h$ and $\epsilon$, the ground state has the lowest energy, thus it is selected by the condition: $\tilde{e}_0={\rm min}_{k_m} e$. The ground state is characterised by its filling value, $k_m$ and the staggered magnetization, $\tilde{x}$, which is calculated self-consistently. In the following we calculate the minimal values of $e(k_m)$ in the two regimes separately, and then comparing those we select $\tilde{e}_0$. To get information about the behavior of $e(k_m)$ we calculate its first two derivatives:
\begin{align}
\frac{{\rm d} e}{{\rm d}k_m}&=\frac{4}{\pi}\left[\frac{h}{2}- \sqrt{J^2 \cos^2 k_m+\epsilon^2 \tilde{x}^2} \right]\;, \\ 
\frac{{\rm d^2} e}{{\rm d}k_m^2}&=\frac{2}{\pi}\frac{J^2 \sin(2k_m)-\epsilon^2 \frac{{\rm d \tilde{x}^2} }{{\rm d }k_m}}{\sqrt{J^2 \cos^2 k_m+\epsilon^2 \tilde{x}^2}}\;.
\label{second_deriv}
\end{align}
The value of the first derivative at the reference points: $k_m=0,\tilde{k}_m$ and $\pi/2$ are given by:
\begin{align}
\begin{split}
\left.\frac{{\rm d} e_0}{{\rm d}k_m}\right|_{0}&=\frac{4}{\pi}\left[\frac{h}{2}- J\right]\;,\\
\left.\frac{{\rm d} e_0}{{\rm d}k_m}\right|_{\tilde{k}_m}&=\frac{4}{\pi}\left[\frac{h}{2}- \frac{J}{\cosh(\pi J/2\epsilon)}\right]\;, \\
\left.\frac{{\rm d} e_0}{{\rm d}k_m}\right|_{\pi/2}&=\frac{4}{\pi}\left[\frac{h}{2}- \epsilon \tilde{x} \right]\;.
\label{e_deriv}
\end{split}
\end{align}
\underline{In the first regime} with $\tilde{x}=0$, $e(k_m)$ is a concave function, since $\frac{{\rm d^2} e}{{\rm d}k_m^2}>0$. The minimal value of $e(k_m)$ is at $k_m=0$, if $\frac{{\rm d} e}{{\rm d}k_m}|_0>0$, thus $h/J>2$. If $\frac{{\rm d} e}{{\rm d}k_m}|_0<0$, but at the same time $\frac{{\rm d} e}{{\rm d}k_m}|_{\tilde{k}_m}>0$ then the minimal value is located in the interior of the first regime. Finally, if $\frac{{\rm d} e}{{\rm d}k_m}|_{\tilde{k}_m}<0$ the minimum of $e(k_m)$ in the first regime is at $\tilde{k}_m$.

\underline{In the second regime} with $\tilde{x}>0$, $e(k_m)$ is a convex function, since $\frac{{\rm d^2} e}{{\rm d}k_m^2}<0$. This can be shown by differentiating the two sides of Eq.(\ref{Self_c}) which leads to the relations:
\begin{align}
\begin{split}
\epsilon^2 \frac{{\rm d \tilde{x}^2} }{{\rm d }k_m}&=\frac{2 (J^2\cos^2 k_m+\epsilon^2 x^2)^{-1/2}}{\int_0^{k_m} {\rm d} k \frac{1}{(J^2\cos^2 k+\epsilon^2 x^2)^{3/2}}}\\
&> \frac{2}{\int_0^{k_m} {\rm d} k \frac{1}{(J^2\cos^2 k+\epsilon^2 x^2)}}> \frac{2 J^2}{\int_0^{k_m} {\rm d} k \cos^{-2} k}\\
&=J^2 \sin (2k_m)/\sin^2 k_m > J^2 \sin (2k_m)\;.
\end{split}
\end{align}
Putting this into Eq.(\ref{second_deriv}) we obtain the announced relation.
In the second regime the minimum of $e(k_m)$ can only be at the boundaries, either at $\tilde{k}_m$ or at $\pi/2$. The $k_m$-dependence of the energy is shown in Fig.\ref{fig_1} at different values of $\varepsilon/J$. Varying the parameter $h/J$ we explore the different phases and phase transitions.

The absolute minimum of $e(k_m)$ can not be at $k_m=\tilde{k}_m$, since it is an inflexion point, thus the possible ground states are of three types.
\begin{itemize}
\item Ferromagnetic (F) ground state with $k_m=0$ and $\tilde{x}=0$. Here there is a finite energy gap, $\Delta E>0$,
and the spin-spin correlation function is zero.
\item Anti-ferromagnetic (AF) ground state with $k_m=\pi/2$ and $1>\tilde{x}>0$. Here the energy gap is a finite, $\Delta E>0$, and the spin-spin correlation function decays exponentially. This can be shown exactly for the end-to-end correlation function using the expression in Eq.(\ref{G1L2}).

\begin{figure}[th!]
\begin{center}
\includegraphics[width=9.cm,angle=0]{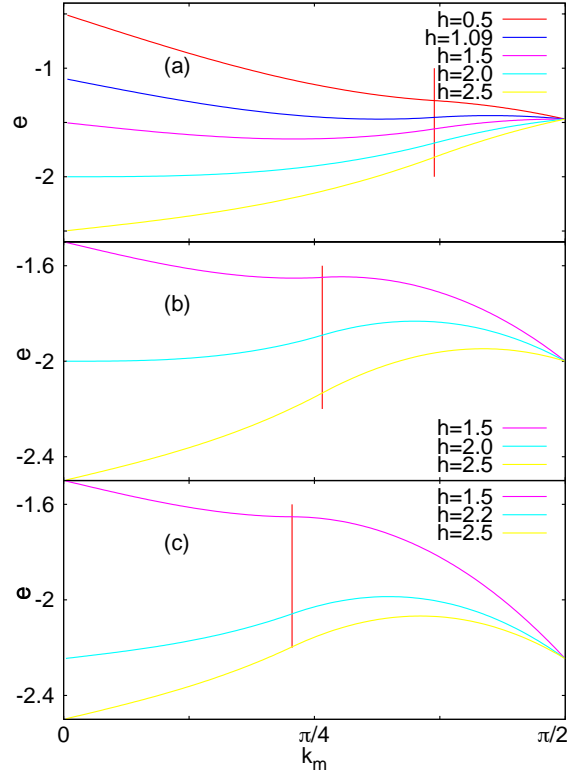} 
\end{center}
\vskip -.5cm
\caption{\label{fig_1} The $k_m$-dependence of the energy at different points of the phase-diagram. The vertical red line shows the position of $\tilde{k}_m$. For $k<\tilde{k}_m$ ($k>\tilde{k}_m$) the staggered magnetization is $x=0$ ($x>0$).
Upper panel - $\varepsilon/J=1$ - : $h/J=0.5$ - AF ground state; $h/J=1.09$ - coexistence between the XY and AF ground states; $h/J=1.5$ - XY ground state; $h/J=2.$ continuous transition from the XY to the F ground state; $h/J=2.5$ - F ground state. Middle panel - $\varepsilon/J=1.7145738$ - : $h/J=1.5$ - AF ground state; $h/J=2.$ - tricritical point, coexistence between the F, XY and AF ground states; $h/J=2.5$ - F ground state. Lower panel - $\varepsilon/J=2.$ - : $h/J=1.5$ - AF ground state; $h/J=2.2$ - coexistence between the F and AF ground states; $h/J=2.5$ - F ground state.}
\end{figure}

\item XY ground state with $0 < k_m< \tilde{k}_m$ and $\tilde{x}=0$. This is a critical ground state, by changing the parameters the filling parameter, $k_m$ is continuously changing. The energy gap is vanishing, $\Delta E=0$, and the spin-spin correlation function decays algebraically. The end-to-end correlations in Eq.(\ref{G1L2}) decays as $G_{1,L} \sim 1/L$.
\end {itemize}
We note that in our model no ground state with simultaneous 
XY and AFM order -- corresponding to supersolid order in the
equivalent hard core BH model --
is realised, which would have $\tilde{k}_m<k_m<\pi/2$ and $\tilde{x}>0$.

\begin{figure}[th!]
\begin{center}
\includegraphics[width=8.5cm,angle=0]{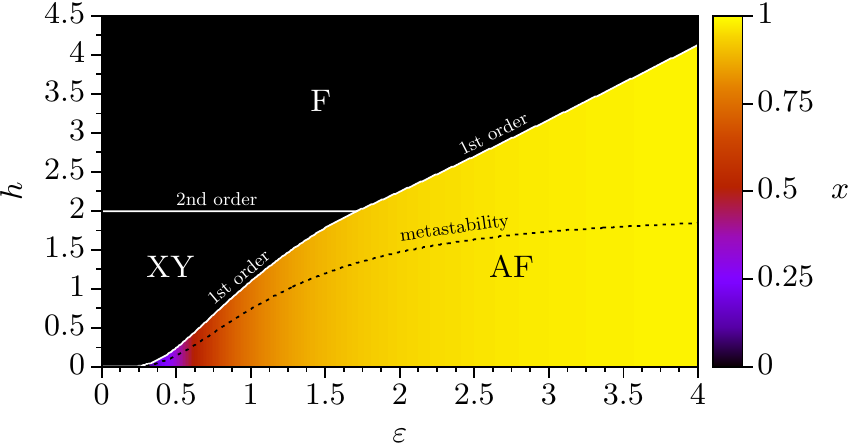}
\includegraphics[width=8.5cm,angle=0]{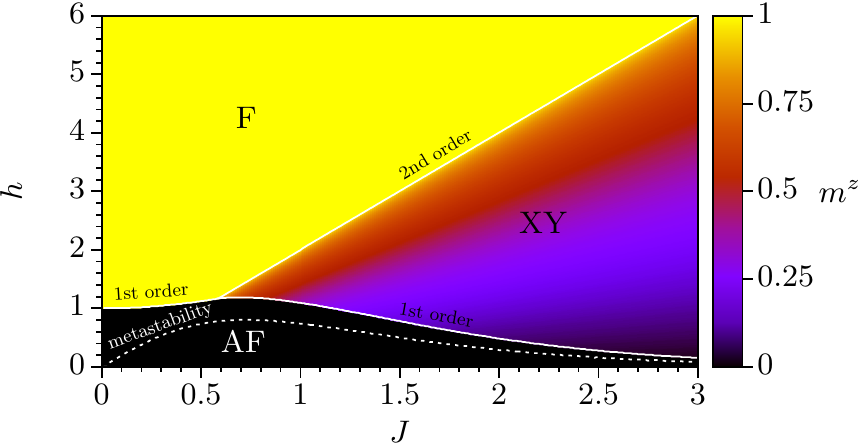}
\end{center}
\vskip -.5cm
\caption{\label{fig_2} Phase-diagram of the quantum XX-model with cavity induced global-range interactions. In the upper panel $J=1$, in the lower panel $\varepsilon=1$. The colour-codes indicate the value of the staggered magnetization, $x$, and that of the longitudinal magnetization, $m^z$.}
\end{figure}

The phase transition between the AF state and the F or the XY state is of first order, there is a jump in the value of $k_m$ at the transition point. On the other hand transition between the XY and the F states is continuous. The phase diagram is shown in Fig.\ref{fig_2}. The metastability limit of the XY-phase is given by: $\left.\frac{{\rm d} e_0}{{\rm d}k_m}\right|_{\tilde{k}_m}=0$, which corresponds according to Eq.(\ref{e_deriv}) $2J/h=\cosh(\pi J/2\epsilon)$.

The AF-phase for small $\varepsilon/J$ is extremly narrow in $h/J$ as can be seen in the upper panel of Fig.\ref{fig_2}. Its extension can be estimated from the analytical form of the metastability limit as $h/J \sim \exp(-\pi J/2 \varepsilon)$. An estimate for the staggered magnetization follows by requiring ${\rm d}e_0/{\rm d}k_m=0$, which from the last equation of (\ref{e_deriv}) gives $\varepsilon x/J \sim \exp(-\pi J/2 \varepsilon)$. This means that the jump of the staggered magnetization at the AF $\to$ XY transition is extremely small for small $\varepsilon/J$ and it goes to zero in a special, exponential form.

\subsection{Spin-spin correlation function}
\label{sec:spin-spin}

We have calculated the spin-spin correlation function, defined as
\be
G(j_1,j_2)=\langle \sigma_{j_1}^x \sigma_{j_2}^x \rangle\;,
\label{G_j1j2}
\ee
which in the free-fermionic description is given by a determinant of order $min(j_2-j_1,L-j_2+j_1)$. The simplest form is given for end-to-end correlations, i.e. with $j_1=1$ and $j_2=L$, and for free chains. In this case with $L=even$ sites the calculation is performed in Appendix C, which leads to the expression:
\begin{align}
\begin{split}
|G(1,L)|=\frac{4}{L+1}\sum_{n=1}^{\frac{k_m}{\pi}L} &\left[1+\left(\frac{\varepsilon x}{J \cos \frac{n \pi}{L+1}}\right)^2\right]^{-1/2}\\
&\times \sin^2\left( \frac{n \pi}{L+1}\right) (-1)^n \;.
\label{G1L2}
\end{split}
\end{align}
Evidently in the ferromagnetic phase with $k_m=0$ we have $G(1,L)=0$, which follows also from symmetry. In the XY-phase with $\tilde{x}=0$ and $0<k_m<\pi/2$ the sum in Eq.(\ref{G1L2}) with terms of alternating signs will result in an uncompensated term of ${\cal O}(L^{-1})$, thus the end-to-end correlations decay algebraically as 
\begin{align}
G(1,L) \sim \frac{1}{L} ,\quad {\rm XY-phase}\;.
\end{align}
Thus in the XY-phase there is quasi-long-range spin order.

Finally in the AF-phase with $\tilde{x}>0$ and $k_m=\pi/2$ the expression in Eq.(\ref{G1L2}) is similar to the form of sine square deformation\cite{sine_square} and this leads to an exponential correction in $L$, thus the end-to-end correlations decay exponentially:
\begin{align}
G(1,L) \sim \frac{1}{L} \exp\left( - \frac{L}{\xi} \right),\quad {\rm AF-phase}\;.
\end{align}
 We have demonstrated this by evaluating Eq.(\ref{G1L2}) numerically. The correlation length is a monotonously increasing function of $\tilde{x} \equiv \frac{\varepsilon x}{J}$. In the limit $\tilde{x} \ll 1$ it goes like:
\begin{align}
\xi \approx \frac{1}{\tilde{x}},\quad \tilde{x} \ll 1\;.
\end{align}
Here we can use the estimate of $\tilde{x}$ at the end of the previous section, which leads to $\xi \sim \exp(\pi J/2 \varepsilon)$, which grows exponentially, in somewhat similar way as at the Kosterlitz-Thouless transition.

The correlation length can be calculated analytically in the opposite limit $\tilde{x} \gg 1$, when in the r.h.s. of Eq.(\ref{G1L2}) we perform a Taylor expansion:
\begin{align}
\begin{split}
|G(1,L)|=\frac{4}{L+1}\sum_{n=1}^{L/2} &\left[\sum_{k=0}^{\infty}\frac{(2k-1)!!}{(2k)!!} \left( \frac{\cos \frac{n \pi}{L+1}}{\tilde{x}} \right)^{2k}(-1)^k \right]\\
&\times \frac{\cos \frac{n \pi}{L+1}}{\tilde{x}}\sin^2\left( \frac{n \pi}{L+1}\right) (-1)^n \;.
\label{G1L3}
\end{split}
\end{align}
Here using the fact, that:
\begin{align}
\begin{split}
\sum_{n=1}^{L/2} \left(\cos \frac{n \pi}{L+1}\right)^{2k-1} \sin^2\left( \frac{n \pi}{L+1}\right) (-1)^n\\
=\begin{cases}
0,\quad {\rm if}\quad k<L/2\\
\ne 0,\quad {\rm if}\quad k \ge L/2
\end{cases} \;,
\label{G1L4}
\end{split}
\end{align}
we obtain
\begin{align}
\begin{split}
|G(1,L)| \sim \frac{4}{L+1} \tilde{x}^{-(L-1)}, \quad \tilde{x} \gg 1\;,
\label{G1L5}
\end{split}
\end{align}
wich defines a correlation length
\begin{align}
\xi \approx \frac{1}{\ln \tilde{x}},\quad \tilde{x} \gg 1\;.
\end{align}

\subsection{Phase-diagram of the non-linearized Hamiltonian}

\begin{figure}[h!]
\begin{center}
\vskip -1cm
\includegraphics[width=\columnwidth,angle=0]{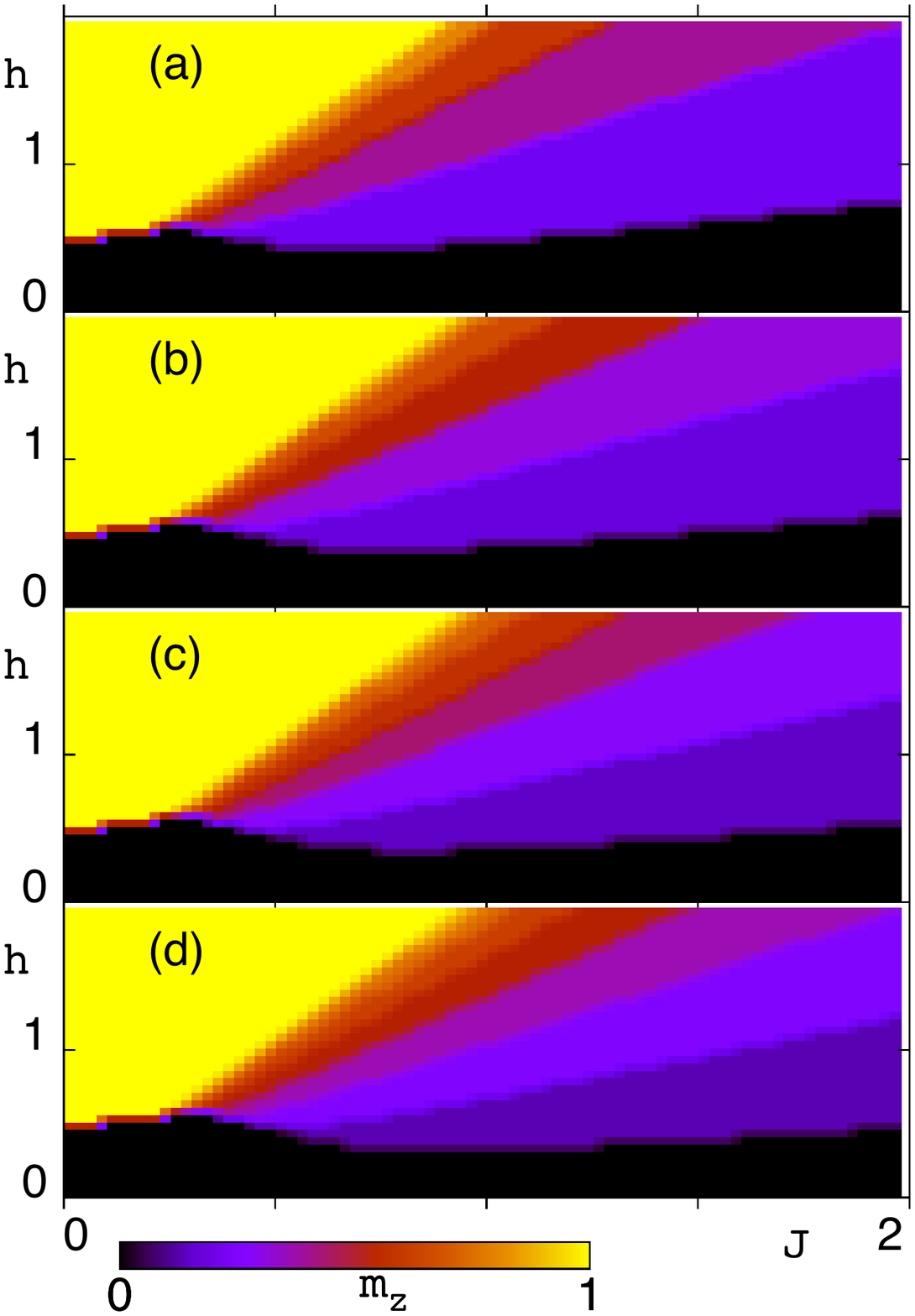}
\end{center}
\vskip -2cm
\caption{\label{fig_3} Phase-diagrams calculated in the Ising basis with the Hamiltonian in Eq.(\ref{XX-Ham1}). From top to bottom $L=10,12,14,16$.}
\end{figure}

We have checked the role of finite-size effects by solving the problem in the original form, given by the Hamiltonian in Eq.(\ref{XX-Ham1}). In particular we have calculated the $z$-component of the magnetization defined by: $m^z=\langle \sum_i \sigma_i^z \rangle/L$, which is shown in Fig.\ref{fig_3} for finite chains with $L=10,12,14$ and $16$. Since $\sum_i \sigma_i^z$ is a conserved quantity, it could have only integer values in the ground-state of the system, therefore in Fig.\ref{fig_3} there are $L/2$ possible discrete values of $0 \le m^z \le 1.$ Also in finite systems the F, XY and AF phases are identified with $m^z=1.$, $0.5 < m^z < 1.$ and $m^z=0.5$, respectively. The finite-size phase-diagrams and the values of the $m^z$ are qualitatively similar for finite values of $L$, as well as in the thermodynamic limit, see in the lower panel of Fig.\ref{fig_2}. Somewhat larger finite-size corrections are present for larger values of $J/\varepsilon$ close to the phase-boundary between the $XY$ and the $AF$ phases.

\vfill

\section{Non-equilibrium dynamics after a quench}
\label{sec:dynamics}

\subsection{Preliminaries}

Let us consider our model defined in Eq.(\ref{XX-Ham1}) in which the parameters are suddenly changed at time $t=0$: from $J_0,h_0,\varepsilon_0$ to $J,h,\varepsilon$. The actual state of the system at $t=0$ is $|\Psi_0 \rangle$, the ground state of the initial Hamiltonian ${\cal H}(J_0,h_0,\varepsilon_0)$ but its time-evolution is governed by the after-quench Hamiltonian, ${\cal H}(J,h,\varepsilon)$ so that the state of the system at time $t>0$ is given by:
\begin{align}
|\Psi(t) \rangle=\exp(-i {\cal H}(J,h,\varepsilon)t) |\Psi_0 \rangle \;.
\end{align}
In our calculation we have used an equivalent Hamiltonian in Eq.(\ref{XX-Ham2}) in which the global-range interaction term is linearised in the thermodynamic limit. This linearised Hamiltonian depends formally on the value of the staggered magnetization, which has to be calculated self-consistently at $t=0$.
For infinitesimal times $\Delta t$ the time evolution operator
$\exp(-i \Delta t{\cal H}(J,h,\varepsilon))$ can formally be treated in the 
thermodynamic limit $L\to\infty$ in the same way as the partition function,
as detailed at the end of Appendix B. Then the saddle point equation in Eq.(\ref{x_k}) holds at each time-step, thus the effective linearized Hamiltonian governing the dynamics assumes the same form as ${\cal H}'$ in Eq.(\ref{XX-Ham2}), however the parameter $x$ is replaced by a time-dependent function $x(t)$, which satisfies the self-consistency criterion:
\begin{align}
x(t)=\langle \Psi(t)|\hat{x} |\Psi(t) \rangle\;,
\label{x(t)}
\end{align}
with
\begin{align}
|\Psi(t) \rangle=\exp\left[-i \int_0^t dt'{\cal H}'({\cal T},\mu,\varepsilon,x(t'))\right] |\Psi_0 \rangle\;.
\end{align}
One can easily show, that the total energy is conserved under the process. Indeed using the Hellmann-Feynmann theorem one obtains:
\begin{align}
\begin{split}
\frac{{\rm d} E_0(t)}{{\rm d} t}&= \left<\Psi(t)\left|\frac{{\rm d} {\cal H'}(t)}{{\rm d} t}\right|\Psi(t) \right>\\
&=-2L\epsilon[\left<\Psi(t)|\hat{x} |\Psi(t) \right>-x(t)]\frac{{\rm d} x(t)}{{\rm d} t}=0\;,
\end{split}
\end{align}
where in the last step the self-consistency equation in Eq.(\ref{x(t)}) is used.

Calculation of time-dependent quantities in the fermionic basis is shown in Appendix D, here we shortly recapitulate the main steps of the derivation. During the quench at $t=0$ new set of free-fermion operators are created, $\gamma_k$ and $\gamma_{k-\pi}$, which are related to the original ones by a rotation, see in Eq.(\ref{G4}). The time-dependent fermion operators, $c_k(t)$ and $c_{k-\pi}(t)$ are expressed with $\gamma_k$ and $\gamma_{k-\pi}$ through time-dependent Bogoliubov parameters in Eq.(\ref{eta2}). These generally complex parameters satisfy a set of differential equations in Eq.(\ref{GG}), which contain $x(t)$ in Eq.(\ref{Stt}) and has to be integrated with the known initial conditions at $t=0$.

\subsection{Numerical results}

The calculation of the state of the system after time $t$ from the quench necessitates the integration of a set of $(L+1)$ linear differential equations of complex variables. This integration has been performed numerically by the fourth-order Runge-Kutta method and the step-size was chosen appropriately to obtain stable results. We have checked by comparing results of non-equilibrium relaxation with sizes $L$ and $2L$, that finite-size effects are negligible until time $t<t^* \sim L$. In the vicinity of non-equilibrium critical points where the time-scale is divergent (see in Eqs.(\ref{minima}) and (\ref{scaling})) we went up to $L=8192$.

We note, that the term with the transverse field: ${H}_{tr}=h\sum_j\sigma_j^z$ commutes with the Hamiltonian: $\left[{H}',{H}_{tr}\right]=0$, therefore the wavefunction of a given state of ${H}'$ does not depend on $h$ (but its energy naturally does). Consequently a quench from $h_0$ to $h \ne h_0$ with fixed $\epsilon_0/J_0=\epsilon/J$ does not modify the stationary properties of the system. In the following we concentrate on those quenches in which both $h$ and $J$ are kept fixed and only the parameter of the global-range interaction term changes from $\epsilon_0$ to $\epsilon$. 

\subsection{Quench from the AF phase}

\begin{figure}[h!]
\vskip -.5cm
\begin{center}
\includegraphics[width=8.5cm,angle=0]{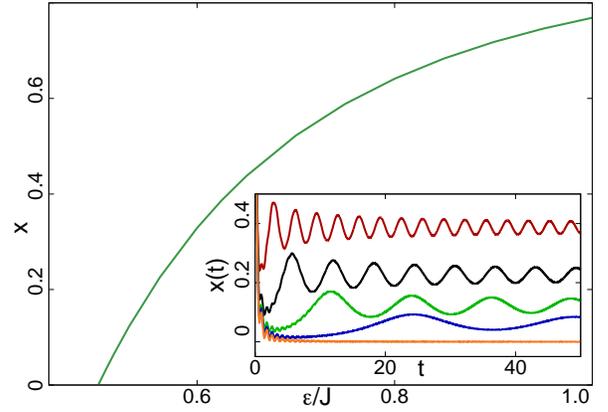}
\end{center}
\vskip -.5cm
\caption{\label{fig_4} Inset: Time-dependence of the staggered magnetization after a quench protocol with $\epsilon_0/J_0=1$ to $\epsilon/J=(\epsilon/J)_c(1+\delta)$ with $(\epsilon/J)_c=1/2$ and $\delta=0.25,~0.125,~0.0625,~0.03125$ and $0$, from up to down. Main panel: Stationary value of the staggered magnetization after a quench from the AF phase with $\epsilon_0/J_0=1$ to $\epsilon/J<\epsilon_0/J_0$. There is a the dynamical phase-transition at $(\epsilon/J)_c=1/2$, at which point  $x_{st}\simeq 1.06 \delta$.}
\end{figure}

In this subsection the initial state we consider belongs to the ordered AF phase, thus $k_m=\pi/2$ and the staggered magnetization at $t=0$ is given by $x(0)>0$. First we consider the case when $\epsilon_0< \infty$, thus the initial state is not fully antiferromagnetic, $1>x(0)>0$. As an example we choose $\epsilon_0/J_0=1$, fix $h_0/J_0=h/J=1$ and vary $\epsilon/J$ for the final state. The time-dependence of the staggered magnetization for different values of the reduced control parameter $\delta=(\epsilon/J)/(\epsilon/J)_c-1$ with $(\epsilon/J)_c=1/2$ are  shown in the inset of Fig.\ref{fig_4}. 

If the strength of the global range interaction term is reduced, $\epsilon/J<\epsilon_0/J_0$, the staggered magnetization shows a fast decay, and after this initial period, characterised by the time and value of the absolute minima, $t_{min}$ and $x_{min}$, respectively, $x(t)$ has an oscillatory behaviour and after sufficiently long time it attains a stationary value $x_{st}$. As a general trend $x_{st}$ is monotonously decreasing with $\epsilon/J$, and for too strong quenches, $\epsilon/J \le (\epsilon/J)_c$, thus $\delta \le 0$, $x_{st}$ vanishes, thus the system exhibits a non-equilibrium dynamical phase transition. The variation of $x_{st}$ with $\epsilon/J$ is shown in Fig.\ref{fig_4}: it vanishes linearly at the phase-transition point:
\be
x_{st} \sim \delta\;.
\label{beta}
\ee

We have observed similar behaviour of $x(t)$ for different initial AF states, the non-equilibrium dynamical phase-transition is found numerically to satisfy the relation:
\be
(\epsilon/J)_c=\frac{\epsilon_0/J_0}{1+\epsilon_0/J_0}\;.
\ee

\begin{figure}[h!]
\vskip -.5cm
\begin{center}
\includegraphics[width=\columnwidth]{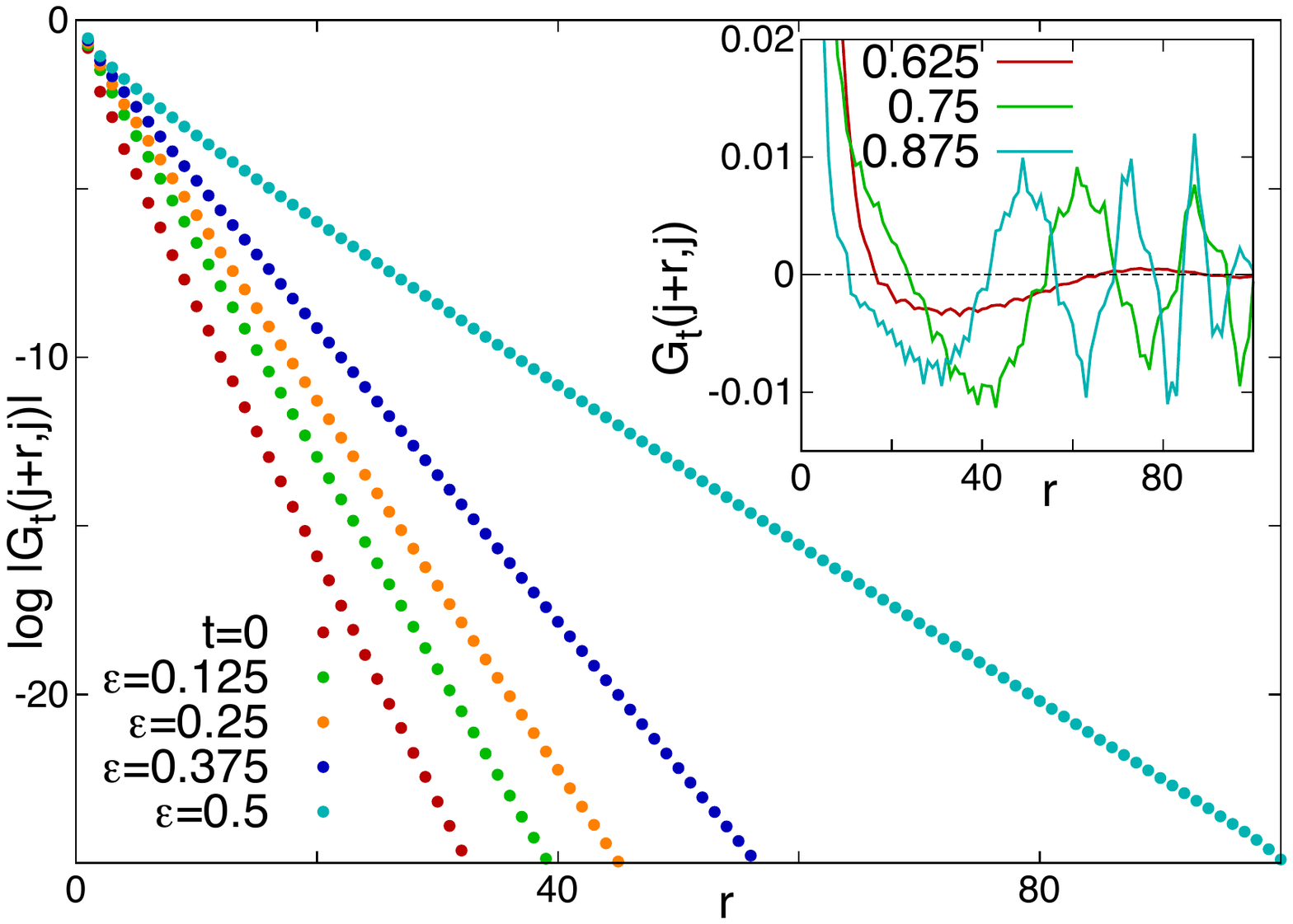}
\end{center}
\vskip -4cm
\caption{\label{fig_5} Non-equilibrium spin-spin correlation functions after a quench protocol with $\epsilon_0/J_0=1$ to $\epsilon/J= 0.125,0.25,0.375$ and $0.5$, at time $t=20$, compared with the equilibrium value at $t=0$. The decays are exponential and the correlation length increases with $\epsilon/J$. Inset: the same for quenches to the dynamically generated AF phase with $\epsilon/J= 0.625,0.75$ and $0.875$. Notice, that $G_t(j+r,j)$ changes sign and for odd or even distance between the spins there is an alternation, which is a sign of the dynamical AF order.}
\end{figure}

We have calculated the non-equilibrium spin-spin correlation function: $G_t(j+r,j)=\langle \sigma^x_{j+r}\sigma^x_{j}\rangle_t$ after the quench at time $t$. In the AF phase using periodic chains the equilibrium spin-spin correlation function: $G_0(j+r,j)$ has an exponential $r$-dependence, similarly to the end-to-end correlation function in Sec.\ref{sec:spin-spin}. This is illustrated in Fig.\ref{fig_5}. If the quench is performed to the regime with no dynamically generated AF order, i.e. with $\epsilon/J \le (\epsilon/J)_c$, then for sufficiently long time $G_t(j+r,j)$ approaches a stationary behavior with an exponential decay, see in Fig.\ref{fig_5}. The correlation length increases with $\epsilon/J$.

The behavior of $G_t(j+r,j)$ changes, if the quench is performed to the region of dynamically generated AF order, i.e. for $\epsilon/J > (\epsilon/J)_c$. In this case, as illustrated in the inset of Fig.\ref{fig_5} $G_t(j+r,j)$ changes sign and has an oscillatory $r$-dependence.

\begin{figure}[h!]
\begin{center}
\includegraphics[width=\columnwidth]{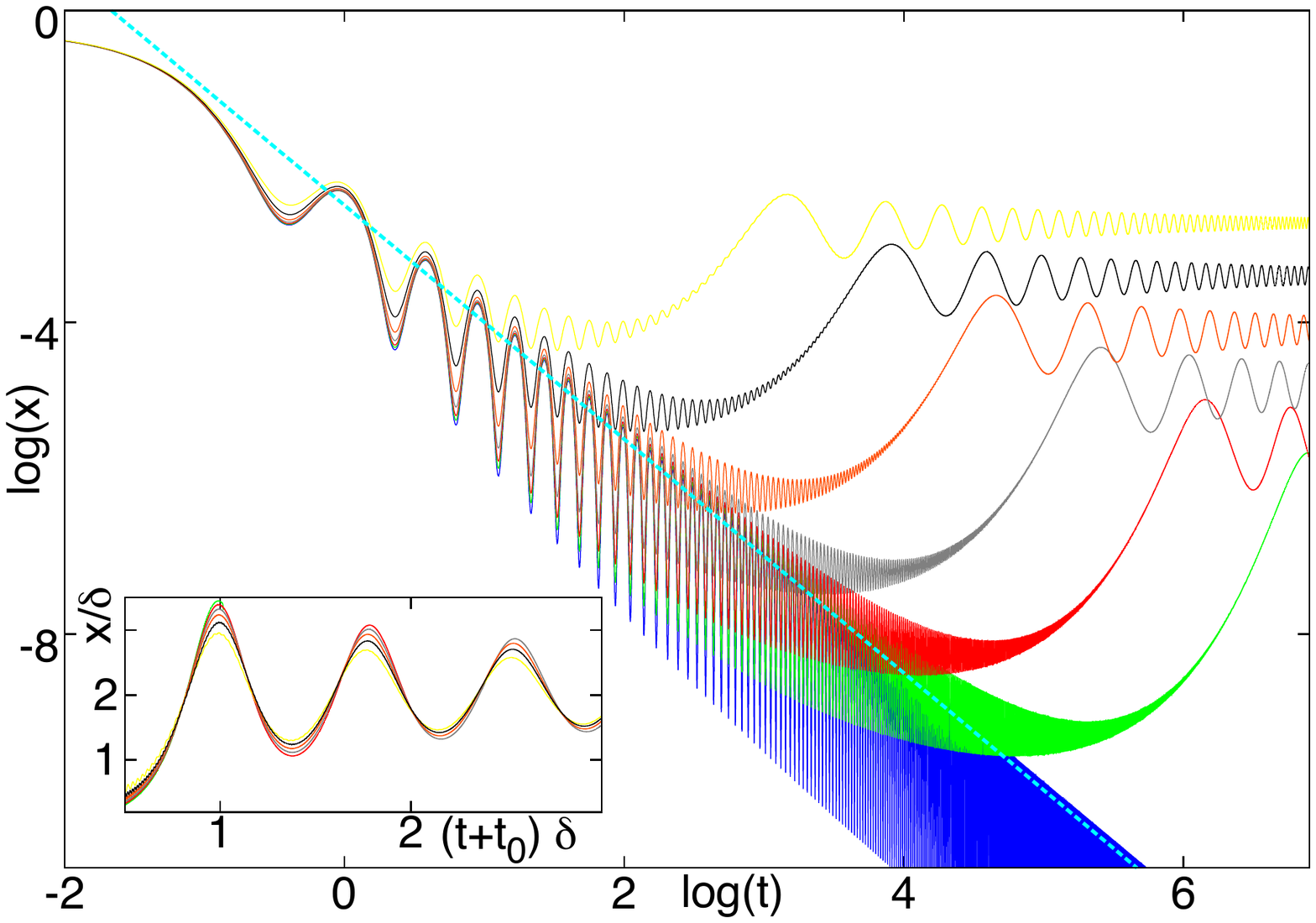}
\end{center}
\vskip -4cm
\caption{\label{fig_6} Time-dependence of the staggered magnetization after a quench protocol: $\epsilon_0/J_0=1.$ and $\epsilon/J=(\epsilon/J)_c(1+\delta)$, with $(\epsilon/J)_c=1/2$ and $\delta=0.,~0.001,~0.002,~0.004,~0.008,~0.016,~0.0032$ from down to up in log-log scale. The straight line with slope $-3/2$ indicates the asymptotic behaviour at the critical point, see in Eq.(\ref{sigma}). Inset: scaling plot of the staggered magnetization curves in the main panel using the relations in Eqs.(\ref{minima}) and (\ref{scaling}).}
\end{figure}

In the following we keep $\epsilon_0/J_0=1$ and concentrate on the behaviour of $x(t)$ in the vicinity of the non-equilibrium dynamical phase transition. The numerical results for $x(t)$ vs. $t$ in log-log scale are collected in Fig.\ref{fig_6}.

At the critical point the staggered magnetization for large time decays algebraically in an oscillatory fashion:
\be
x(t) \sim t^{-\sigma}\sin(\omega_0t),\quad \delta=0 \;.
\label{sigma}
\ee
Our numerical results indicate $\sigma \approx 3/2$ and $\omega_0 \approx 8J$, see in Fig.\ref{fig_6}.

\begin{figure}[h!]
\begin{center}
\includegraphics[width=\columnwidth]{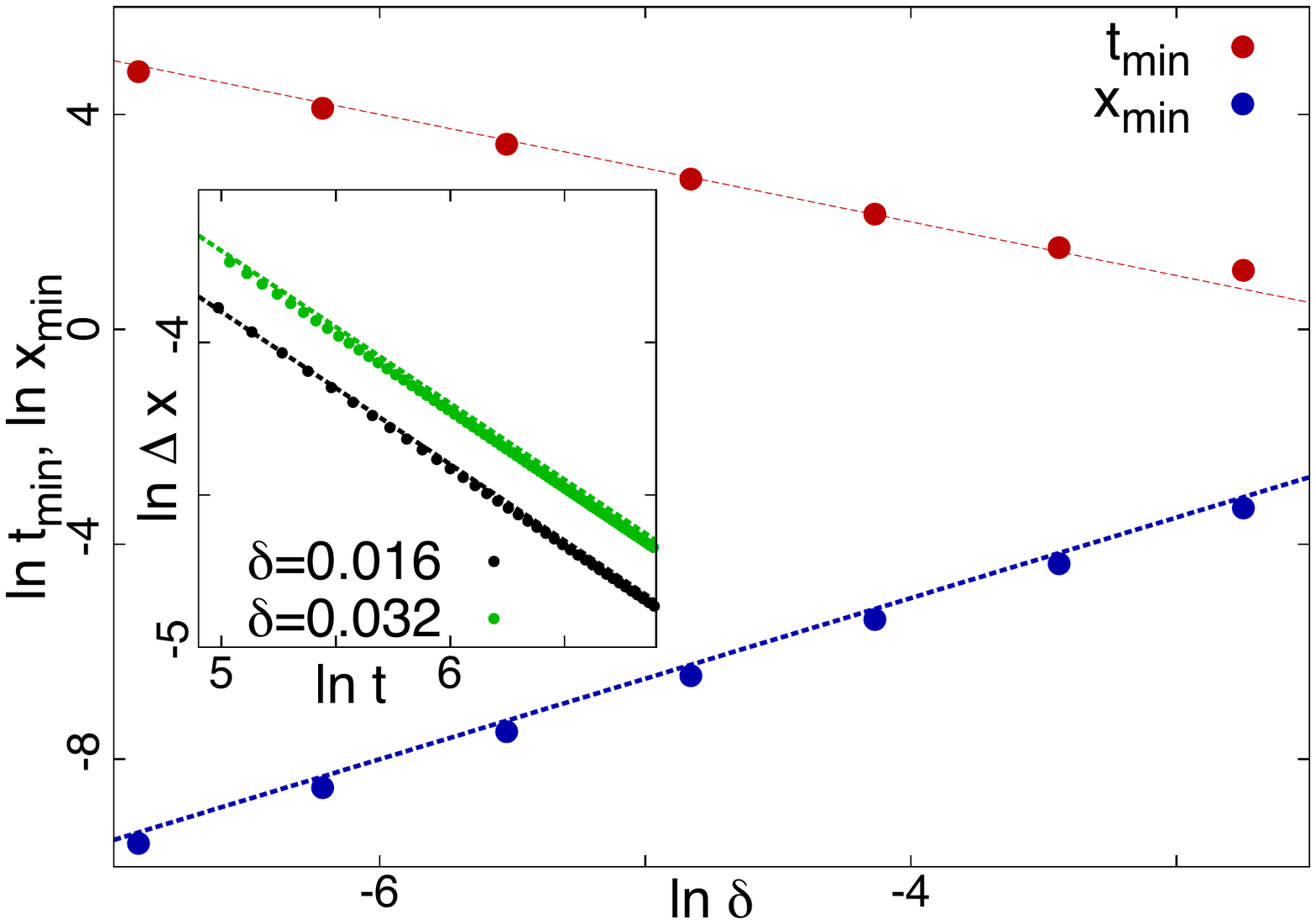} 
\end{center}
\vskip -4cm
\caption{\label{fig_7} Position, $t_{min}$ (upper set of points) and value $x_{min}$ (lower set of points) of the minimum of the dynamical staggered magetization as a function of $\delta$ in log-log scale. The straight lines with slopes $-1$ and $1.5$ indicate the respective expected asymptotic behaviours in Eq.(\ref{minima}). Inset: Amplitude of the oscillations in the stationary region, measured as the difference between two consecutive extremal (maxima and minima) values, as a function of time in log-log scale for $\delta=0.032$ (upper set of points) and $\delta=0.016$ (lower set of points). The straight lines with slope $1/2$ represent the conjectured behavior in Eq.(\ref{ossc}).}
\end{figure}

In the AF ordered regime with $\delta>0$ the $x(t)$ curves start to deviate from the critical one and have a minimum at a characteristic time $t_{min}(\delta)$, having a value $x_{min}$. Close to the transition point these scale as:
\be
t_{min}(\delta) \sim \delta^{-1},\quad x_{min} \sim t_{min}^{-\sigma} \sim \delta^{\sigma}\;,
\label{minima}
\ee
see in Fig.\ref{fig_7}.

After passing the minimum the dynamical staggered magnetization grows to a stationary value and start to oscillate in the form:
\be
x(t) \approx x_{st}(\delta)+\Delta(\delta) t^{-1/2}\sin(\omega(\delta)t),\quad t \gg t_{min}\;,.
\label{ossc}
\ee
in which the amplitude of the oscillations goes to zero as $t^{-1/2}$, which is illustrated in the inset of Fig.\ref{fig_7}.

The time-scale in the stationary region, $\tau(\delta)=1/\omega(\delta)$ is in the same order of magnitude as $t_{min}$, thus there is just one time-scale in the problem, which is divergent at the dynamical phase-transition point:
\be
\tau(\delta)=1/\omega(\delta) \sim \delta^{-1}, \quad \epsilon_0<\infty\;.
\label{scaling}
\ee
Since the prefactor, $\Delta(\delta)$ has only a weak $\delta$ dependence close to the critical point the curves in the main Fig.\ref{fig_6} could be scaled together, which is shown in the inset.

We note that the staggered magnetization after passing the absolute minima has a strong revival, since the ratio of its values at the final (stationary) period and the initial (minimum) period is given by : $x_{st}/x_{min} \sim \delta^{-1/2}$, which is divergent as the transition point is approached.

We have checked, that the values of the scaling exponents are universal for $\epsilon_0<\infty$. If, however the starting state is fully antiferromagnetic, i.e. $\epsilon_0=\infty$ and thus $x(0)=1$, than the appropriate scaling combinations in the stationary regime are the following:
\be
x_{st}(\delta) \sim \delta^{1/2}, \quad \omega(\delta) \sim \delta^{1/2}, \quad \epsilon_0=\infty\;,
\label{eps0_infty}
\ee
which can be illustrated by an appropriate scaling plot of the $x(t)$ curves (not shown here). This difference is due to the fact, that for $\epsilon_0=\infty$ the Bogoliubov-parameters are symmetric: $g_{k,k}(t)=g_{k-\pi,k}(t)$ and $g_{k,k-\pi}(t)=g_{k-\pi,k-\pi}(t)$, which is not the case for $\epsilon_0<\infty$.

\subsection{Quench from the XY phase}

\begin{figure}[h!]
\begin{center}
\includegraphics[width=\columnwidth]{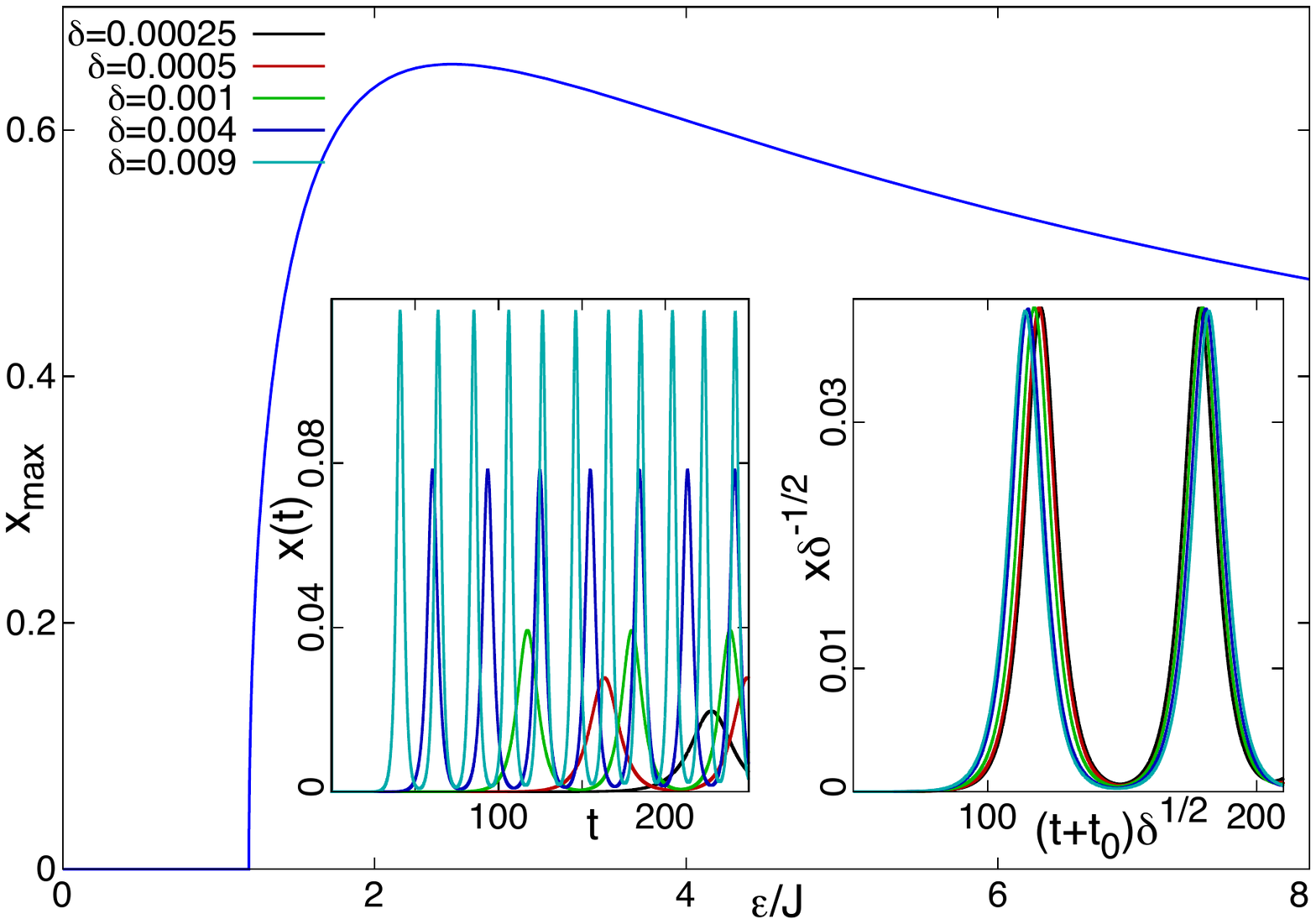}
\end{center}
\vskip -4cm
\caption{\label{fig_8} Right inset: Time-dependence of the staggered magnetization after a quench from the XY-phase from $\epsilon_0/J_0=.5$ with a perturbation $\Delta x=10^{-10}$ to different values of the reduced control-parameter $\delta=0.009,~0.004,~0.001,~0.0005$ and $0.00025$, from up to down. Left inset: Scaling plot using Eq.(\ref{x_max,t_max}) and keeping the plot of $\delta=0.001$ unscaled. Main panel: $\epsilon/J$-dependence of $x_{max}$.}
\end{figure}

In this subsection the initial state belongs to the XY-phase thus we have $0 < k_m< \tilde{k}_m$ and $x(0)=0$. If $x(0)$ is exactly $0$, then according to Eqs.(\ref{GG}) 
the Bogoliubov parameters decouple from each other and $x(t)$ stays zero for $t>0$. In the following we test the stability of this solution by adding a small perturbation, $\Delta x$, to the staggered magnetization. Having fixed $h_0/J_0=h=J=1$ and $\epsilon_0/J_0=0.5$ we have quenched the system to various values of $\epsilon/J$. For smaller values of $\epsilon/J<(\epsilon/J)_c(h) \simeq 1.1946$ the resulting staggered magnetization oscillates around the mean value of $\overline{x}(t)=0$ and its amplitude stays in the order of $\Delta x$. If, however, we quench the system to $\epsilon/J>(\epsilon/J)_c(h)$, then $x(t)$ grows within a time $t_{max}$ to a maximum value of $x_{max}=O(1)$ and then oscillates between $x_{max}$ and $x_{min}$ with a period $t_{per} \sim t_{max}$. This is illustrated in the right inset of Fig.\ref{fig_8}. We note, that similar type of macroscopic revival of an order-parameter has been observed also in Ref.\cite{Schuetz_Morigi}.

This process represents a dynamical phase-transition separating a region in which the solution $x(t)=0$ is stable from a region, in which the time-average value of the dynamically generated staggered magnetization is finite. The dynamical phase-transition coincides with the metastability line in the AF phase, thus dynamically generated staggered magnetization takes place only in such regions, in which no metastable XY-type solution exist.

\begin{figure}[]
\begin{center}
\includegraphics[width=0.75\columnwidth,angle=0]{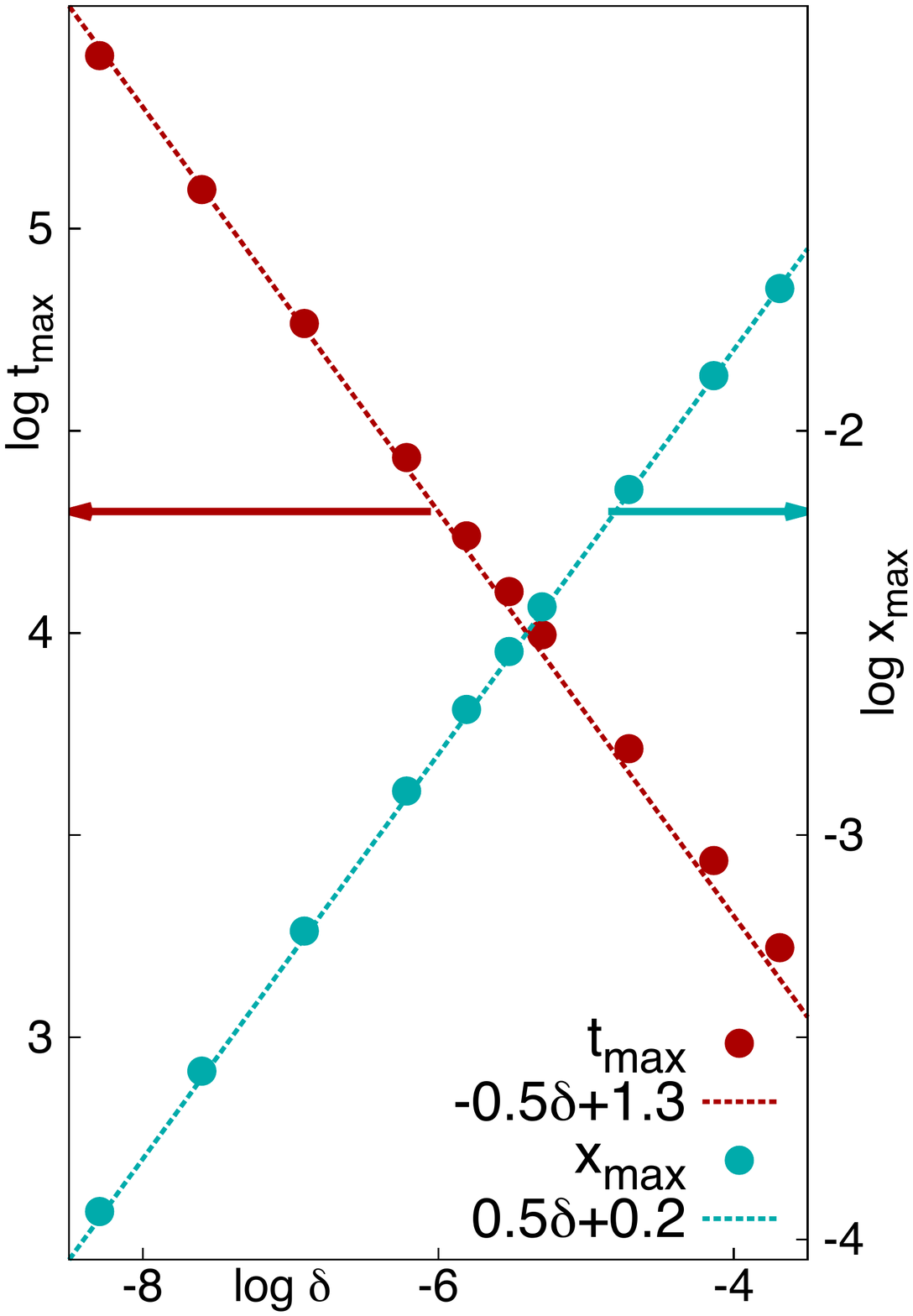}
\end{center}
\vskip-1.5cm
\caption{\label{fig_9} $\delta$-dependence of the (first) maximum $x_{max}$ (right scale) and the time-scale, $t_{max}$ (left scale) in Fig.\ref{fig_8} in log-log scale. The straight lines with slopes $0.5$ and $-0.5$, respectively, represent the expected asymptotic relations in Eq.(\ref{x_scaling}).}
\end{figure}

Denoting the reduced control-parameter by $\delta=(\epsilon/J)/(\epsilon/J)_c - 1$ the maximum value of the dynamically generated staggered magnetization vanishes as a power of $\delta$, but at the same time the time-scale, $t_{max}$, is divergent. Our numerical results in Fig.\ref{fig_9} are consistent with the asymptotic relations:
\be
x_{max} \sim \delta^{1/2}, \quad t_{max} \sim \delta^{-1/2}, \delta \ll 1 \;.
\label{x_max,t_max}
\ee
Close to the dynamical phase-transition point the dynamical staggered magnetization follows the scaling form:
\be
x(t,\delta)=\delta^{1/2} \tilde{x}(t \delta^{1/2}), \delta \ll 1\;,
\label{x_scaling}
\ee
which is illustrated in the left inset of Fig.\ref{fig_8}. Consequently the time-average value of the staggered magnetization for small $\delta$ behaves as:
\beqn
\overline{x}(\delta)&=&\frac{1}{t_{max}}\int_0^{t_{max}} x(t,\delta) {\rm d} t \nonumber \\
&=& \delta^{1/2} \frac{1}{t'_{max}}\int_0^{t'_{max}} \tilde{x}(t') {\rm d} t' \sim \delta^{1/2}\;,
\eeqn
with $t'=t \delta^{1/2}$.

For quenches more deep into the AF phase $x_{max}(\delta)$ goes over a maximum and then decays for large-$\delta$ as $x_{max} \sim \delta^{1/2}$, but we still have $t_{max} \sim \delta^{-1/2}$, as shown in Fig. \ref{fig_9}.

\begin{figure}
\begin{center}
\vskip-6cm
\includegraphics[width=\columnwidth,angle=0]{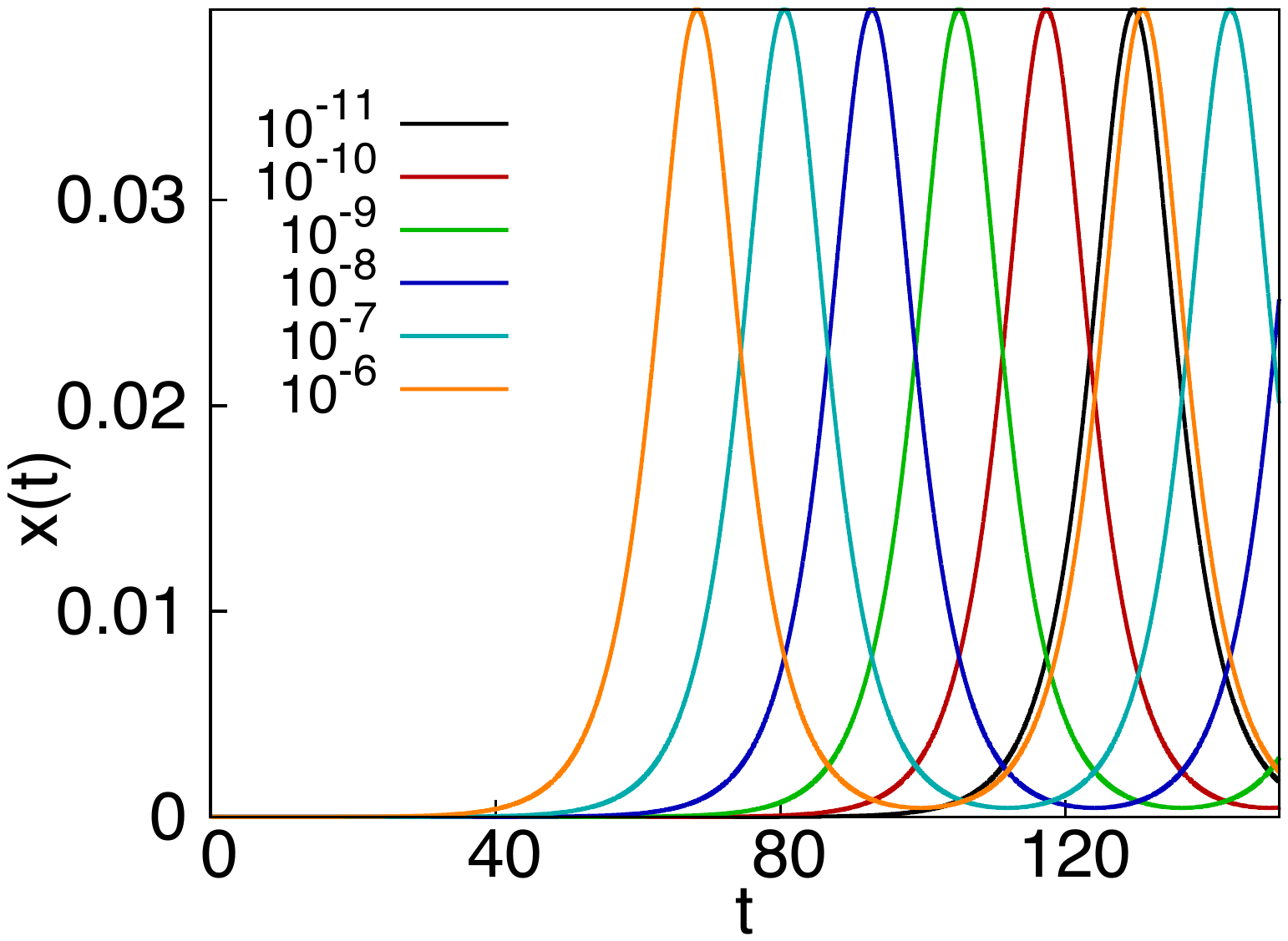}
\end{center}
\vskip -2cm
\caption{\label{fig_10} Quench from the XY-phase with $h_0=h=1$, $J_0=J=1$, $\epsilon_0=0.5$ and $\delta=0.001$ for different values of the perturbation $\Delta x=10^{-6}, 10^{-7},\dots 10^{-11}$, from left to right.}
\end{figure}

We have also checked the effect of the strength of the perturbation $\Delta x$ on the relaxation process. As shown in Fig.\ref{fig_10} with decreasing $\Delta x$ the $x(t)$ curves are simply shifted in time, thus $x_{max}$ stays the same but  the time-scales are shifted by an amount of
\be
t_{shift} \sim \log \Delta x\;.
\ee
Consequently any non-zero perturbation causes a measurable increase of the dynamical staggered magnetization if the quench is performed to $\epsilon/J>(\epsilon/J)_c(h)$.

\begin{figure}[t]
\vskip-3.5cm
\begin{center}
\includegraphics[width=\columnwidth]{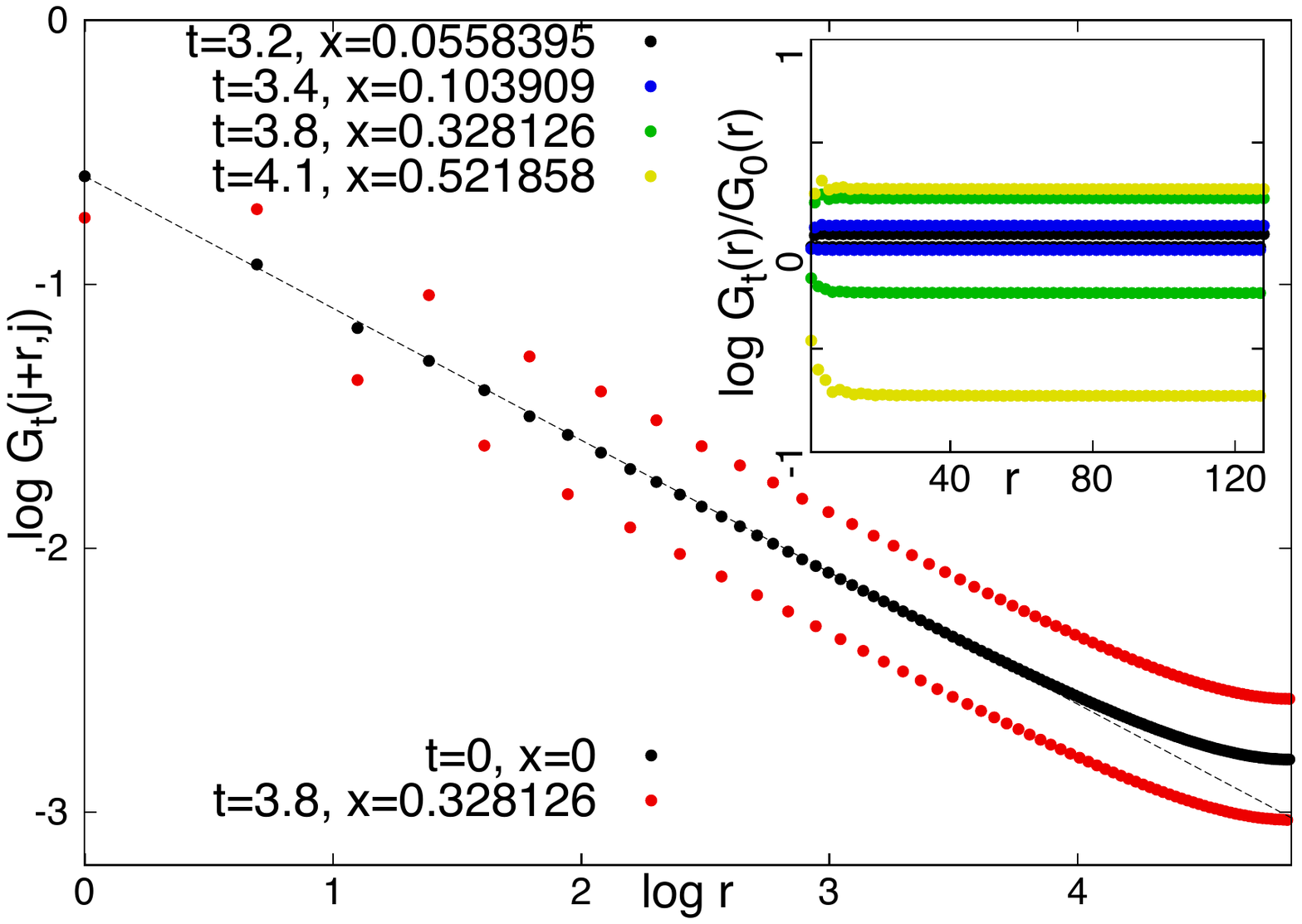}
\end{center}
\vskip -4cm
\caption{\label{fig_11} The equilibrium spin-spin correlation function $G_0(j+r,j)$ and its non-equilibrium counterpart $G_t(j+r,j)$ after a quench protocol from $\epsilon_0/J_0=0.5$ to $\epsilon/J= 1.5$ at $t=3.8$ and $x(t)=0.328$. The straight line with slope $-1/2$ indicates the exact asymptotic behaviour. The algebraic decay of $G_0(j+r,j)$ is preserved under the dynamical phase transition, but for $x(t)>0$ the prefactor is different for even and odd distances between the spins. Inset: Ratio of  $G_t(j+r,j)$ and  $G_0(j+r,j)$ at different times after a quench protocol from $\epsilon_0/J_0=0.5$ to $\epsilon/J= 1.5$. The difference between the ratios of different parity is larger for larger values of $x(t)$.}
\end{figure}

For periodic boundary conditions we have measured the non-equilibrium spin-spin correlation function $G_t(j+r,j)$ after a quench protocol from $\epsilon_0/J_0=0.5$ to different values of $\epsilon/J$. The equilibrium spin-spin correlations in the XY phase are found to show an algebraic decay, similarly to the end-to end correlations in Sec.\ref{sec:spin-spin}, and the value of the decay exponent is consistent with the exact asymptotic relation: $G_0(j+r,j) \sim r^{-1/2}$, see in Fig.\ref{fig_11}. If the quench is performed below the dynamical phase-transition point $\epsilon/J \le (\epsilon/J)_c$ where $x(t)=0$ the non-equilibrium spin-spin correlation function is identical with $G_0(j+r,j)$. If, however after the quench there is a dynamically generated AF order, i.e. $\epsilon/J > (\epsilon/J)_c$ and $x(t)>0$, then the shape of $G_t(j+r,j)$ is different from $G_0(j+r,j)$. The algebraic decay is preserved, but for $x(t)>0$ the prefactor is different for even and odd distances between the spins. This is shown in Fig.\ref{fig_11}. We have also calculated the ratio $G_t(j+r,j)/G_0(j+r,j)$ at different times after the quench, which is shown in the inset of Fig.\ref{fig_11}. This ratio is different for even and odd distances between the spins but practically independent of the value of $r$ of the given parity. The larger the order-parameter, $x(t)$, the larger the difference between the ratios.

We can thus conclude that after a quench from the XY-phase to the AF-phase above the metastability line such a state is created, in which AF order and XY quasi-long-range order coexist. The analogous state in the Bose-Hubbard model is the supersolid phase.

\section{Discussion}
\label{sec:discussion}

In this paper we have considered the quantum XX-model in the presence of a transverse field and with competing short- and long-range interactions. This type of system has been realised experimentally by ultracold atoms in optical lattices and a global-range interaction term is mediated by the presence of a high-finesse optical cavity\cite{Science-Esslinger,Klinder2015,Landig2016}. 
Here we have considered the experimental set up when the lattice constant of the optical lattice is half the wave-length of the cavity mode and the global-range interaction term is expressed as the square of the staggered magnetization. In the thermodynamic limit the global-range interaction term can be linearised, so that the expectation value of the staggered magnetisation is obtained through a self-consistent treatment, like in mean-field models.

The one-dimensional problem is transformed to a fermionic model which has been solved exactly by standard techniques. The ground-state phase-diagram of the model consists of three phases. The XY-phase, in which the spin-spin correlations decay algebraically, persists for moderately strong transverse fields and global-range interactions. For strong transverse fields the system is ferromagnetic, while for strong global-range interactions the system is antiferromagnetic with a non-vanishing staggered magnetization. In equilibrium there is no such ground state, in which XY- and antiferromagnetic order are present at the same time, which corresponds to the supersolid phase in the equivalent hard-core BH model.

We have also considered the non-equilibrium, quench dynamic of the system, which is also accessible experimentally. We have shown, that the linearization of the global-range interaction term can be performed in this case too, such that the staggered magnetization has to be calculated self-consistently at each time-step. In the quench process the initial and the final states are characterised according to the equilibrium phase diagrams. In a quench from an AF state to an XY state the dynamical staggered magnetization is shown to approach a stationary value, which is finite above a dynamical phase-transition point. In the vicinity of the transition point the order-parameter vanishes continuously and at the same time the characteristic time-scale is divergent. The critical exponents at the non-equilibrium transition are universal, i.e. independent of the initial state, except when the initial state is fully antiferromagnetic. In the quench in the opposite direction, i.e. from an initial XY-state to an AF final state we have studied the stability of the time-dependent staggered magnetization by adding a small perturbation to the trivial solution $x(t)=0$. In this process also a dynamical phase transition is observed, the transition point of which coincides with the metastability line of the XY-state. This point separates the regime in which the solution $x(t)=0$ is stable from that, in which the time-average of the staggered magnetization is finite. In the latter regime dynamically generated AF order exists on top of the XY-quasi-long-range order, which is analogous to the supersolid state of the BH model.

The method of solution presented in this paper is applicable for a set of one-dimensional spin, hard-core boson or fermion models with competing short- and long-range interactions. We expect that also these systems exhibit a rich equilibrium phase-diagram and interesting non-equilibrium quench dynamics.

\begin{figure}[t]
\begin{center}
\vskip-3.7cm
\includegraphics[width=\columnwidth,angle=0]{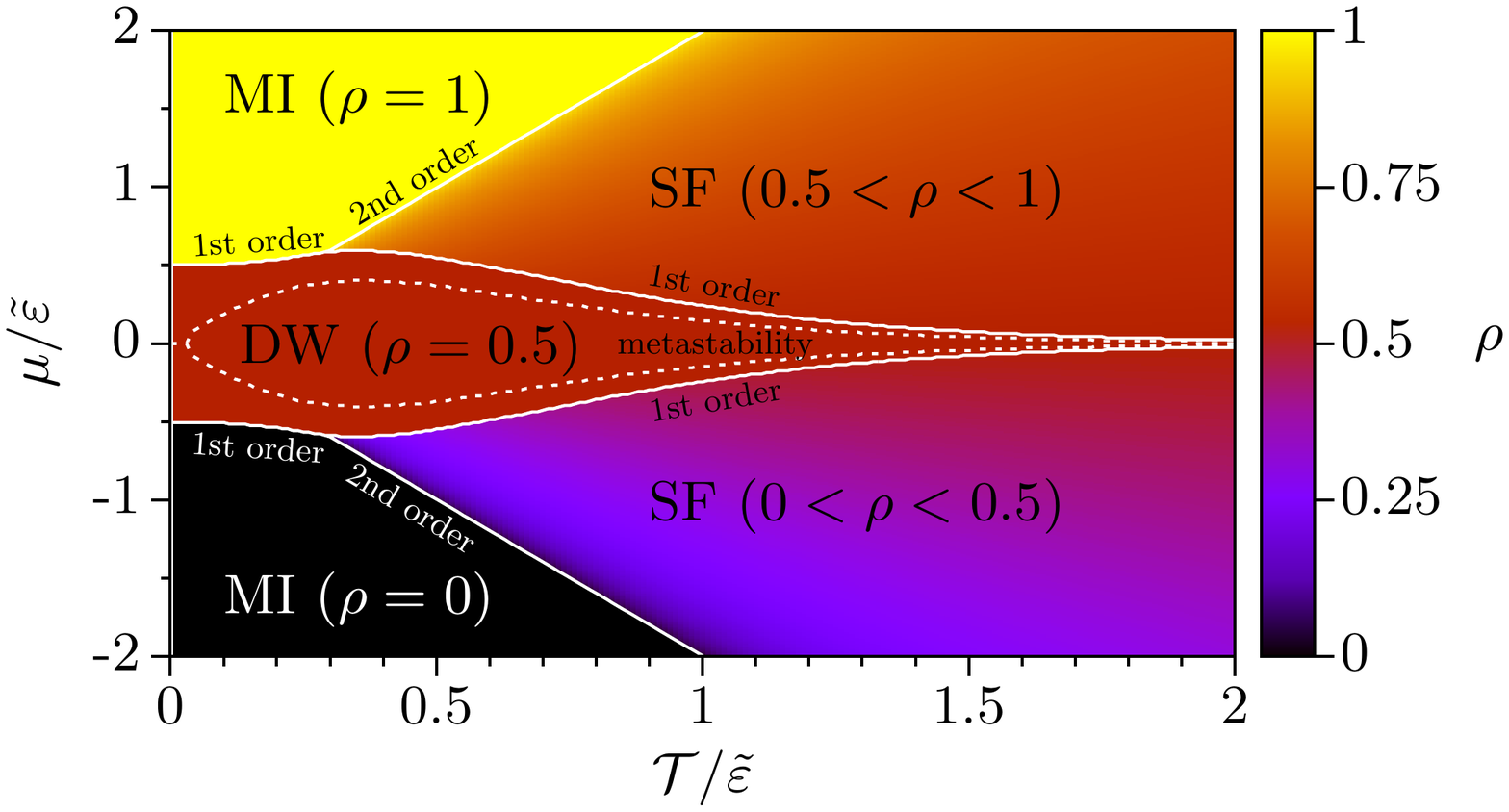}
\end{center}
\vskip-4.5cm
\caption{\label{fig_12} Phase diagram of the one-dimensional Bose-Hubbard model with cavity mediated 
global range interaction in the hard core limit obtained by using the equivalence
with the XX-model analyzed in this work. $\rho$ is particle density and $x$ the 
occupancy imbalance. MI = Mott insulator, SF = superfluid, DW = density wave.}
\end{figure}

Our model is equivalent to a one-dimensional extended BH model of hard-core bosons with equilibrium phases: Mott insulating (MI), superfluid (SF) and density wave (DW), which correspond to the F, XY and AF phases, respectively. 
Utilizing this equivalence we analyzed in \cite{Blass-PRL} the one-dimensional 
Bose-Hubbard model with cavity mediated global range interaction in the hard
core limit. The corresponding $\mu-{\cal T}$ phase diagram is shown in Fig.\ref{fig_12}, which is 
simply the phase diagram displayed in Fig.\ref{fig_2} translated to the Bose-Hubbard
nomenclature and extended to negative chemical potentials (longitudinal fields 
in the magnetic context). 

It should be noted that the superfluid (SF) phase is actually, in 1d, one with 
quasi-long range order signaled by the algebraic decay of the SF correlation function $\tilde{G}(r)=\langle b_{i+r}^\dagger b_{i}\rangle$. This is related to the spin-spin correlation function in the $XX$ model, see in Eq.(\ref{G_j1j2}),
via the relations displayed in Eq.(\ref{bose-spin}) and the xy-isotropy of the Hamiltonian (\ref{XX-Ham1}). The structure of the phase diagram shares some features with the $\mu-{\cal T}$ phase diagrams of soft core Bose-Hubbard models with cavity mediated global range interactions \cite{Dogra2016,Flottat2017}: the $\rho=0$ and $\rho=1$ MI regions are reminiscent of the corresponding Mott lobes embedded in the SF region in the soft core system, sandwiching between them a DW lobe with $\rho=1/2$. An important difference to the soft-core system is the absence of a super-solid region (with simultaneous SF and DW order) 
at the tip of the DW lobe, which instead extends in a thin protrusion to arbitrarily large hopping strengths at $\mu=0$.

For quenches from the SF phase into the DW phase across the metastability line we find that the SF correlation functions still decay algebraically indicating
the simultaneous presence of quasi-long-range SF order and (time-averaged) 
DW order, see section V.D. Consequently the system attains dynamically
generated supersolid (SS) properties after strong enough quenches from 
the DW into the SF phase, which is not the ground state, but 
a high energy state. Furthermore it is interesting to note that for times with $x(t)>0$ there is an even-odd modulation of the SF correlation functions which increases with $x(t)$ and which disappears when the imbalance goes back to $0$
(see Fig. 5 in \cite{Blass-PRL}. The density modulations reflect even-odd modulations of the SF correlation functions.

Thus we predict that (time-averaged) SS properties emerge during the 
time evolution of a SF initial state in a Bose-Hubbard system 
under the influence of sufficiently strong cavity mediated long-range interactions. Superfluidity is not lost and the periodically modulated site occupation imbalance builds up beyond a critical interaction strength.
The dynamical emergence of diagonal long-range DW order on top of 
off-diagonal (quasi)-long-range SF order is a feature of the
non-equilibrium dynamics of closed quantum system that has 
to our knowledge never been reported before. Since its origin 
is the presence of the global range interactions we expect it to be 
robust and to be observable also in two- and three-dimensional 
Bose-Hubbard systems with cavity-induced interactions, for hard-core
as well as soft-core bosons. It would be interesting to check these
predictions with, for instance, tVMC mehods \cite{tVMC}.

It should also be emphasized that in 1d the ground state 
phase diagram does not display a SS region (Fig. 1). 
Here we have shown that high energy states 
(not eigenstates of the Hamiltonian) can dynamically generate 
SS order with periodically modulated DW and SF correlations
in the stationary state. Since superfluidity is destroyed
at finite temperature in 1d the stationary high-energy state with SS 
properties that we find cannot be described by a finite temperature
equilibrium ensemble. Consequently the system we analyzed does not 
thermalize for some quenches, which is particularly remarkable 
considering the fact that for finite size (finite $L$) the system 
is non-integrable (it is integrable only for $L\to\infty$).

Our predictions of remanent, metastable DW order after DW$\to$SF quenches
and the dynamical generation of periodically modulated DW order 
superposed to metastable SF order after SF$\to$DW quenches can 
be tested experimentally in a cavity-setup like 
the one used in \cite{Landig2016}, even though this setup is 
two-dimensional and involves soft-core bosons. Preliminary 
experimental indications for such metastability phenomena occurring after 
quenches of the cavity induced interaction strength have indeed
been reported recently \cite{Esslinger-Metastab}. 

For an even quantitative experimental reproduction of our exact results 
one would have to modify the setup used in \cite{Landig2016}
to establish an ensemble of 1d optical lattices in the deep lattice limit
as in \cite{Haller-etal} and to record the time evolution 
as e.g. in \cite{Kaufmann-etal}. Other experimental constraints are:
1) the presence of a harmonic trap generating a wedding cake like 
organization of SF, MI and DW regions, which is straightforward to 
include into our model calculations by a spatially 
varying chemical potential; 2) a finite experimental system size, 
which will not play a role 
for times smaller than a scale set by the inverse maximum group velocity \cite{Igloi-Rieger}; 3) different experimental quench protocols, which
can also be straightforwardly into our analysis; and 4) the cavity photon 
loss of that might influence the dynamics on long time scales 
by causing decoherence -- which, however, can be discarded on short 
time scales depending on the loss rate \cite{BH-Cavity1,BH-Cavity2,Schuetz_Morigi}.
Independent of theses details our exact results will serve as
a firm reference for the interpretation and understanding 
of quench-experiments with lattice bosons with cavity mediated 
long-range interactions.

\section*{Appendix A: Extended Bose-Hubbard model}

In the experimental setup Rb atoms are placed on a optical lattice which is prepared inside an ultrahigh-finesse optical cavity and the lattice constant of the optical lattice is half the wave length of the cavity mode\cite{Landig2016}. Theoretically this many-body system of bosons is described by an extended Bose-Hubbard (BH) model
\cite{BH-Cavity1,BH-Cavity2} the Hamiltonian of which is given for a bipartite lattice:
\begin{align}
\begin{split}
\hat{H}_{BH}=&-\mathcal{T}\sum_{\braket{\mathbf{r},\mathbf{r}'}}\left(\hat{b}_{\mathbf{r}}^{\dagger}\hat{b}_{\mathbf{r}'}+\text{H.c.}\right)+\frac{U}{2}\sum_{\mathbf{r}}\hat{n}_{\mathbf{r}}\left(\hat{n}_{\mathbf{r}}-1\right)
\\
&-\mu\sum_{\mathbf{r}}\hat{n}_{\mathbf{r}}
-\tilde{\varepsilon}\frac{1}{N}\left(\sum_{\mathbf{r}\in e}\hat{n}_{\mathbf{r}}-\sum_{\mathbf{r}\in o}\hat{n}_{\mathbf{r}}\right)^2
\label{BH-Ham}
\end{split}
\end{align}
where $\hat{b}_{\mathbf{r}}^{\dagger}$ ($\hat{b}_{\mathbf{r}}$) 
are the Bose creation (annihilation) operators, 
$\hat{n}_{\mathbf{r}}=\hat{b}_{\mathbf{r}}^{\dagger}\hat{b}_{\mathbf{r}}$ the number operators, $N$ the lattice size, 
$\mathcal{T}$ the tunneling constant, $U$ the on-site repulsion, $\mu$ the chemical potential and $\tilde{\varepsilon}$ the strength of the infinite-range interactions induced by the cavity. The cavity-induced long-range 
interactions are represented as the square of the density wave 
order parameter $\hat{x}$
\begin{align}
\hat{x}=\frac{1}{N}\left(\sum_{\mathbf{r}\in e}\hat{n}_{\mathbf{r}}-\sum_{\mathbf{r}\in o}\hat{n}_{\mathbf{r}}\right)
\end{align}
where $e$ and $o$ stand for even and odd lattice sizes, respectively.

In the large-$U$ limit, when multiple occupancy of lattice sites is excluded $\hat{b}_{\mathbf{r}}^{\dagger}$ and $\hat{b}_{\mathbf{r}}$ are replaced by hard-core Bose operators, which can be represented by the Pauli matrices, $\sigma_{\mathbf{r}}^{x,y,z}$ in the following way:
\begin{align}
\begin{split}
\hat{b}_{\mathbf{r}}^{\dagger} &\longrightarrow \frac{1}{2}\left( \sigma_{\mathbf{r}}^{x} + i \sigma_{\mathbf{r}}^{y}\right)\\
\hat{b}_{\mathbf{r}} &\longrightarrow \frac{1}{2}\left( \sigma_{\mathbf{r}}^{x} - i \sigma_{\mathbf{r}}^{y}\right)\\
\hat{n}_{\mathbf{r}} &\longrightarrow \frac{1}{2}\left( 1+ \sigma_{\mathbf{r}}^{z}\right)
\end{split}
\label{bose-spin}
\end{align}
In this representation the BH Hamiltonian in Eq.(\ref{BH-Ham}) is replaced by the Hamiltonian of the quantum XX model:
\begin{align}
\begin{split}
\hat{H}_{XX}=&-J\sum_{\braket{\mathbf{r},\mathbf{r}'}}\left(\sigma_{\mathbf{r}}^{x} \sigma_{\mathbf{r}'}^{x}+\sigma_{\mathbf{r}}^{y}\sigma_{\mathbf{r}'}^{y}\right)\\
&-h\sum_{\mathbf{r}}\sigma^z_{\mathbf{r}}
-\varepsilon\frac{1}{N}\left(\sum_{\mathbf{r}\in e}\sigma^z_{\mathbf{r}}-\sum_{\mathbf{r}\in o}\sigma^z_{\mathbf{r}}\right)^2
\label{XX-Ham}
\end{split}
\end{align}
having the correspondences: $\mathcal{T} \to 2J$, $\mu \to 2h$ and $\tilde{\varepsilon} \to 4 \varepsilon$.

\section*{Appendix B: Linearization of the global-range interaction term}

Consider the Hamiltonian of the XX-chain in Eq.(\ref{XX-Ham1}) and for convenience define the kinetic energy part as
\begin{equation}
\hat{\cal T}=
-J\sum_{i=1}^L (\sigma_i^x \sigma_{i+1}^x + \sigma_i^y \sigma_{i+1}^y )
\end{equation}
and the potential energy part as
\begin{equation}
\hat{\cal V}=
-h\sum_{i=1}^L \sigma_i^z 
-\epsilon L \left( \frac1L\sum_{i,odd}\sigma_i^z - \frac1L\sum_{i,even}\sigma_i^z \right)^2\;.
\end{equation}
Using the Suzuki-Trotter decomposition the canonical partition function can be written as
\begin{equation}
Z={\rm Tr}\;e^{-\beta H}
=\lim_{M\to\infty} {\rm Tr} \left( 
e^{-\Delta\tau\hat{\cal T}} e^{-\Delta\tau\hat{\cal V}} \right)^M\;,
\end{equation}
where $\Delta\tau=\beta/M$ and ground state properties are obtained in the limit $\beta \to \infty$.

We use $\sigma_i^z$ eigenstates, denoted as $|S_1,\ldots,S_L\rangle$
with $S_i=\pm1$ and insert between 
any two factors a representation of unity, $\sum_{\underline{S}^k}|\underline{S}^k\rangle
\langle\underline{S}^k|$ to obtain the 
Feynman path-integral expression for the partition function
\begin{align}
\begin{split}
Z&=\lim_{M\to\infty}\sum_{\underline{S}^1,\ldots,\underline{S}^{M}}
\prod_{k=1}^M
\langle\underline{S}^k|e^{-\Delta\tau\hat{\cal T}}|\underline{S}^{k+1}\rangle\\
&\times \exp\left(-\Delta\tau\sum_{k=1}^M V(\underline{S}^k)\right)\;,
\end{split}
\end{align}
where $\underline{S}^k = (S_1^k,\ldots,S_L^k$) for 
$k=1,\ldots,M$ and we have used that $\hat{\cal V}$ is diagonal in the $\sigma^z$-basis.

The quadratic part in $V(\underline{S}^k)$ can now be decoupled using the identity (Hubbard-Stratonovic transformation):
\begin{equation}
e^{\lambda A^2} = \int\frac{dx}{{\cal N}}\,e^{-\lambda x^2 +2\lambda x A}\;,
\end{equation}
where ${\cal N}$ is a normalization factor. One obtains
\begin{align}
\begin{split}
&\exp\left(-\Delta\tau\sum_{k=1}^M V(\underline{S}^k)\right)\\
&=
\int\prod_{k=1}^M \frac{dx_k}{{\cal N}}\,
\exp\left(\Delta\tau\sum_{k=1}^M \biggl\{ h\sum_{i=1}^L S_i^k\right.\\
&+\left. \epsilon L\biggl[-x_k^2 + 
2x_k\biggl(\frac1L\sum_{i,odd}S_i^k-\frac1L\sum_{i,even}S_i^k\biggr)\biggr]\biggr\}
\right)
\end{split}
\end{align}
The partition function then reads
\begin{align}
\begin{split}
&Z=\lim_{M\to\infty}\sum_{\underline{S}^1,\ldots,\underline{S}^{M}}
\prod_{k=1}^M
T_{k,k+1}
\int\prod_{k=1}^M \frac{dx_k}{{\cal N}}\, \\
&\exp\left(-L\biggl\{
\Delta\tau\sum_{k=1}^M 
[\epsilon x_k^2 -2\epsilon x_k {\cal D}(\underline{S}^k)
-h m(\underline{S}^k)]\right)\;,
\end{split}
\end{align}
with
\begin{eqnarray}
T_{k,k+1}&=&\langle\underline{S}^k|e^{-\Delta\tau\hat{\cal T}}|\underline{S}^{k+1}\rangle\nonumber \\
{\cal D}(\underline{S}^k)&=&\frac1L\sum_{i,odd}S_i^k-\frac1L\sum_{i,even}S_i^k \nonumber \\
m(\underline{S}^k)&=&\frac1L\sum_{i=1}^L S_i^k\;.
\end{eqnarray}

By performing the sum over spins first we can rewrite the 
partition function as 
\begin{eqnarray}
Z=&&\lim_{M\to\infty}
\int\prod_{k=1}^M \frac{dx_k}{{\cal N}}\,\nonumber\\
&&\exp\biggl(-L\biggl\{\Delta\tau\sum_{k=1}^M \left[\epsilon x_k^2 + f_L(\underline{x},\epsilon,h,J)\right]\biggr\}\biggr) 
\label{ZX}
\end{eqnarray}
with $f_L(\underline{x},\epsilon,h,J)$ the free energy 
of a 1+1-dimensional world line model (derived from an XX Hamiltonian)
with Gaussian fluctuating fields coupled to the staggered magnetization
${\cal D}(\underline{S}^k)$ in each of the M (imaginary) time slices:
\begin{align}
\begin{split}
&f_L(\underline{x},\epsilon,h,J)=\\
&-\frac1L\ln\left(
\sum_{\underline{S}^1,\ldots,\underline{S}^{M}}
\prod_{k=1}^M
T_{k,k+1}
\cdot
e^{\Delta\tau\left[2\epsilon x_k {\cal D}(\underline{S}^k)+h m(\underline{S}^k)\right]}\right)\;.
\end{split}
\label{fff}
\end{align}
The free energy
of the original model (\ref{XX-Ham1}) in the 
thermodynamic limit is
$f(\epsilon,h,J)=\lim_{L\to\infty} -\beta^{-1}\ln\,Z$.
Applying the saddle-point method to (\ref{ZX}) in the limit $L\to\infty$
yields
\begin{equation}
f(\epsilon,h,J)={\rm min}_{\underline{x}}\;
\underbrace{
\biggl\{\Delta\tau\sum_{k=1}^M \left[\epsilon x_k^2 + f_L(\underline{x},\epsilon,h,J)\right]\biggr\}}_{=:g(\underline{x})}\;.
\end{equation}
The saddle point equation $\partial g/\partial x_k = 0$ then read
\begin{equation}
x_k=\langle {\cal D}(\underline{S}^k) \rangle\;.
\label{x_k}
\end{equation}
where $\langle\ldots\rangle$ denotes the $\underline{x}$-dependent 
thermal average.

Since the Hamiltonian is time-independent the observables have to
be (imaginary) time-translational invariant, too. Thus $x_k$ is 
independent of the Trotter slice index $k$, i.e. $x_k=x$
$\forall k=1,\ldots,M$ with $x=\langle{\cal D}(\underline{S}^1)\rangle$.

Consequently in the thermodynamic limit $L\to\infty$ the partition function
is given by 
\begin{equation}
{\cal Z}={\rm Tr}\,\exp(-\beta {H}'(x))
\end{equation}
with
\begin{equation}
{H}'(x) = L\varepsilon x^2 - 2\varepsilon L x \hat{x} - h\hat{m}
+\hat{\cal T}\,,\label{AppH}
\end{equation}
where $\hat{x}$ and $\hat{m}$ are the staggered and 
longitudinal magnetization, respectively, and
$x$ has to be determined self-consistently via
\begin{equation}
x=\langle\hat{x}\rangle
={\rm Tr}\,\hat{x}\,\exp(-\beta {H}'(x))/{\cal Z}\;.\label{Appx}
\end{equation}
At zero temperature, $\beta\to\infty$ the expectation value 
$\langle\cdots\rangle$ becomes the expectation value in the
ground state of ${H}'(x)$.

Analogously one shows that for infinitesimal times 
$\Delta t$ the matrix elements of the time evolution operator 
${\cal U}(\Delta t)=\exp(-i\Delta t\,{H})$ is given by
\begin{equation}
\langle\underline{S}\vert {\cal U}(\Delta t) \vert\underline{S}'\rangle
=\langle\underline{S}\vert \exp(-i\Delta t\,
{H}'(x)) \vert\underline{S}'\rangle
\end{equation}
with ${H}'(x)$ as in (\ref{AppH}). 
Note that for infinitesimal time steps the corresponding
integral in (\ref{ZX}) is over only one auxiliary variable $x$
and $f_L(x,\varepsilon,h,J)$ in (\ref{fff}) is now
\begin{align}
\begin{split}
&f_L(x,\varepsilon,h,J)=\\
&-\frac{1}{L}\ln\left(
\langle\underline{S}\vert\exp(-i\Delta t\,{\cal T})\vert
\underline{S}'\rangle\cdot 
e^{-i\Delta t [2\varepsilon x {\cal D}(\underline{S})
+hm(\underline{S})]}
\right).
\end{split}
\end{align}
Therefore the saddle point equation for $x$ now simply demands that
$x=\langle\underline{S}\vert \hat{x}\vert\underline{S}'\rangle
\cdot\delta_{\underline{S},\underline{S}'}$,
implying that it is given by the staggered magnetization of 
the state $\underline{S}$.
For the time evolution of a particular state $\vert\psi\rangle$
this implies that in each infinitesimal time step the 
operator ${H}'(x)$ has to be applied to $\vert\psi(t)\rangle$
with an $x$ equal to the staggered magnetization of the state
at time $t$. Thus $x$ becomes, as expected, time dependent
and has to be calculated as described in Appendix D.

\section*{Appendix C: End-to-end correlations in free chains}
\label{AppC}

Here we calculate end-to-end correlations in free chains, the Hamiltonian of which is given by Eq.(\ref{XX-Ham2}), however in the first term of the r.h.s. the sum runs up to $L-1$. Consequently in the fermionic expression in Eq.(\ref{hamiltonian_mf1}) the second term in the r.h.s. is missing. In the following we fix $L=even$. To diagonalise this Hamiltonian we use the canonical transformation:
\begin{align}
c_j=\sum_k\sum_{\pm} \phi^{(\pm)}_k(j) \eta^{(\pm)}_k\;,
\end{align}
so that $\phi^{(\pm)}_k(j)$ are real and these are represented by two standing waves at odd and even sites:
\begin{align}
\begin{split}
\phi^{(\pm)}_k(2l-1)&=a_1^{(\pm)}(k) \sin[k(L+2-2l)]\\
\phi^{(\pm)}_k(2l)&=a_2^{(\pm)}(k) \sin[k(L+1-2l)]\;.
\label{phi}
\end{split}
\end{align}
The wave-numbers are given by: $k=\frac{n}{L+1}\pi$, for $n=1,2,\dots,L/2$, and for each $k$ there are two free-fermionic modes with energy:
\begin{align}
\Lambda_k^{(\pm)}=-2h \pm 4\sqrt{J^2 \cos^2k+\varepsilon^2 x^2}\;.
\end{align}
The corresponding prefactors are: 
\begin{align}
\begin{split}
a_1^{(-)}(k)=a_2^{(+)}(k)=\left[ 1 + \left(\sqrt{a_k^{-2}+1}+a_k^{-1} \right)^2 \right]^{-1/2}\\
a_2^{(-)}(k)=a_1^{(+)}(k)=\left[ 1 + \left(\sqrt{a_k^{-2}+1}-a_k^{-1} \right)^2 \right]^{-1/2}\;,
\label{a12}
\end{split}
\end{align}
with $a_k=\frac{J }{\epsilon x}\cos k$ as in Eq.(\ref{g}).
Note, that with respect to the solution of periodic chains $\Lambda_k^{(-)}$ ($\Lambda_k^{(+)}$) corresponds to $\Lambda_k$ and ($\Lambda_{k-\pi}$) in Eq.(\ref{Lambda}) and the diagonalised Hamiltonian in Eq.(\ref{hamiltonian_mf3}) is valid with the correspondences: $\eta^{(-)}_k \to \eta_k$ and $\eta^{(+)}_k \to \eta_{k-\pi}$. It is easy to check, that both the ground-state energy and the staggered magnetisation assumes equivalent expressions in the thermodynamic limit, as obtained for periodic chains in section \ref{Sec:phase_diagram}.

End-to-end correlations are defined as:
\begin{align}
G(1,L)=\langle \sigma_1^x \sigma_L^x \rangle = \langle \Psi_0 | (c^{\dag}_1 + c_{1})(c^{\dag}_{L} - c_{L}) w | \Psi_0 \rangle\;,
\label{G1L}
\end{align}
where
\begin{align}
| \Psi_0 \rangle = \left[\sum_{0 < k < \pi/2 }\eta^{\dag(-)}_k + \sum_{k_m < k < \pi/2 }\eta^{\dag(+)}_k \right]| 0 \rangle;
\end{align}
is the ground state of the Hamiltonian and $w$ is defined below Eq.(\ref{hamiltonian_mf1}). Substituting Eq.(\ref{phi}) into (\ref{G1L}) we get for the absolute value of the end-to-end correlation function:
\begin{align}
\begin{split}
|G_{1,L}|=\sum_k \left[\phi^{(-)}_k(1)\phi^{(-)}_k(L) \langle \Psi_0 |1 -2 \eta^{\dag(-)}_k\eta^{(-)}_k | \Psi_0 \rangle \right. \\
\left. +\phi^{(+)}_k(1)\phi^{(+)}_k(L) \langle \Psi_0 |1 -2 \eta^{\dag(+)}_k\eta^{(+)}_k | \Psi_0 \rangle \right]\;,
\label{G1L1}
\end{split}
\end{align}
which is evaluated as given in Eq.(\ref{G1L2}).

\section*{Appendix D: Non-equilibrium dynamics in the free-fermion representation}

In the non-equilibrium process we perform a quench at time $t=0$, when the set of parameters in the Hamiltonian in Eq.(\ref{XX-Ham2}) are suddenly changed from $J_0,h_0,\varepsilon_0$ to $J,h,\varepsilon$. During the quench new set of free-fermion operators are created, $\gamma_k$ and $\gamma_{k-\pi}$, which are related to the original set of free-fermion operators, $\eta_k$ and $\eta_{k-\pi}$ in Eq.(\ref{eta}) in the following way:
\begin{align}
\begin{split}
\gamma_k&= \cos \delta_k\eta_k-\sin \delta_k \eta_{k-\pi}\\
\gamma_{k-\pi}&= \sin \delta_k\eta_k+\cos \delta_k \eta_{k-\pi}\;.
\label{G4}
\end{split}
\end{align}
Here $\delta_k=\Theta_k-\Theta_k^{(0)}$ is the difference between the Bogoliubov-angles: $\tan 2 \Theta_k=-\frac{\epsilon x(0)}{J \cos k}$ and $\tan 2 \Theta_k^{(0)}=-\frac{\epsilon_0 x(0)}{J_0 \cos k}$, furthermore $x(0)$ is the solution of the self-consistency equation in Eq.(\ref{SK}) with the Hamiltonian ${\cal H'}(J_0,h_0,\varepsilon_0)$.

The time-dependence of the fermion operators for $t \ge 0$ are given by:
\begin{align}
\begin{split}
c_k(t)&=u_{k,k}(t)\gamma_k+u_{k-\pi,k}(t)\gamma_{k-\pi}\\ 
c_{k-\pi}(t)&=u_{k,k-\pi}(t)\gamma_k+u_{k-\pi,k-\pi}(t)\gamma_{k-\pi} \;,
\label{eta2}
\end{split}
\end{align}
where the Bogoliubov-parameters are generally complex for $t>0$. We note that at $t=0$ these are real and can be written by the Bogoliubov-angles:
\begin{align}
\begin{split}
u_{k,k}(0)&=u_{k-\pi,k-\pi}(0)=\cos \Theta_k, \\
u_{k,k-\pi}(0)&=-u_{k-\pi,k}(0)=-\sin \Theta_k\;,
\label{Theta_k}
\end{split}
\end{align}
furthermore expressing $c_k(0)$ and $c_{k-\pi}(0)$ through $\gamma_k,\gamma_{k-\pi}$ and then through $\eta_k,\eta_{k-\pi}$ leads to the relation in Eq.(\ref{G4}).

Time derivative of the fermion operators, $c_k(t)$ and $c_{k-\pi}(t)$ can be calculated in the Heisenberg picture: ${\rm d}c_{k}(t)/{\rm d} t=i[{\cal H}_k,c_{k}]$ and ${\rm d}c_{k-\pi}(t)/{\rm d} t=i[{\cal H}_k,c_{k-\pi}]$, which are linear in the $c_k$-s, since ${\cal H}_k$ is quadratic in the fermion operators:
\begin{align}
\begin{split}
\frac{{\rm d}}{{\rm d} t}c_{k}&=-i\left[-2(h+2J\cos k)c_{k}+4\epsilon x c_{k-\pi}\right]\\
\frac{{\rm d}}{{\rm d} t}c_{k-\pi}&=-i\left[4\epsilon x c_{k}+-2(h-2J\cos k) c_{k-\pi}\right]\;.
\label{H1}
\end{split}
\end{align}
Similar relations hold for the creation operators, $c^{\dag}_{k}(t)$ and $c^{\dag}_{k-\pi}(t)$.

Inserting now Eq.(\ref{eta2}) into Eq.(\ref{H1}) we obtain a set of differential equations for the Bogoliubov-parameters:
\begin{align}
\begin{split}
\frac{{\rm d}u_{k,k}}{{\rm d} t}&=-i\left[-2(h+2J\cos k)u_{k,k}+4\epsilon x(t) u_{k,k-\pi}\right]\cr
\frac{{\rm d}u_{k-\pi,k}}{{\rm d} t}&=-i\left[-2(h+2J\cos k) u_{k-\pi,k}+4\epsilon x(t) u_{k-\pi,k-\pi}\right]\cr
\frac{{\rm d}u_{k,k-\pi}}{{\rm d} t}&=-i\left[-2(h-2J\cos k)u_{k,k-\pi}+4\epsilon x(t) u_{k,k}\right]\cr
\frac{{\rm d}u_{k-\pi,k-\pi}}{{\rm d} t}&=-i\left[-2(h-2J\cos k) u_{k-\pi,k-\pi}+4\epsilon x(t) u_{k-\pi,k}\right]\;,
\label{GG}
\end{split}
\end{align}
with the initial condition at $t=0$ given in Eq.(\ref{Theta_k}). In Eq.(\ref{GG}) the actual value of the staggered magnetization at time $t$ is given by $x(t)=\langle \hat{x}(t) \rangle$ (see also in Eq.(\ref{x(t)})), where $\hat{x}(t)$ is defined in terms of fermion operators in Eq.(\ref{Eq:Imbalance_c}). This is expressed with the time-dependent Bogoliubov-parameters and the occupation numbers in the initial free-fermionic basis as:
\begin{widetext}
\begin{align}
\begin{split}
x(t)= \langle \hat{x}(t) \rangle&=\frac{2}{L}\sum_{k>0} \left[ (u^*_{k,k}u_{k,k-\pi}+u_{k,k}u^*_{k,k-\pi})\left(\cos^2\delta_k\left< \eta^{\dag}_k \eta_k \right>+\sin^2\delta_k\left< \eta^{\dag}_{k-\pi} \eta_{k-\pi} \right> \right)\right.\cr
&+\left.(u^*_{k-\pi,k}u_{k-\pi,k-\pi}+u_{k-\pi,k}u^*_{k-\pi,k-\pi})\left(\sin^2\delta_k\left< \eta^{\dag}_k \eta_k \right>+\cos^2\delta_k\left< \eta^{\dag}_{k-\pi} \eta_{k-\pi} \right> \right)\right.\cr
&+\left.(u^*_{k,k}u_{k-\pi,k-\pi}+u_{k-\pi,k}u^*_{k,k-\pi})\frac{\sin 2\delta_k }{2}\left(\left< \eta^{\dag}_k \eta_k \right>-\left< \eta^{\dag}_{k-\pi} \eta_{k-\pi} \right> \right)\right.\cr
&+\left.(u^*_{k-\pi,k}u_{k,k-\pi}+u_{k,k}u^*_{k-\pi,k-\pi})\frac{\sin 2\delta_k }{2}\left(\left< \eta^{\dag}_k \eta_k \right>-\left< \eta^{\dag}_{k-\pi} \eta_{k-\pi} \right> \right)\right]\;.
\label{Stt}
\end{split}
\end{align}
\end{widetext}
It is easy to check that $x(t)$ is continuous at $t=0$, as it should be. In the actual calculation one should determine the time-dependence of the Bogoliubov-parameters and the staggered magnetization, which necessities the integration of a set of $(L+1)$ coupled first-order differential equations with complex variables. At some point it is of interest to calculate the time-derivative of the staggered magnetization, which is expressed as:
\begin{widetext}
\begin{align}
\begin{split}
\frac{{\rm d} x(t)}{{\rm d} t}&=i\frac{8J}{L}\sum_{k>0} \cos k\left[ (u^*_{k,k}u_{k,k-\pi}-u_{k,k}u^*_{k,k-\pi})\left(\cos^2\delta_k\left< \eta^{\dag}_k \eta_k \right>+\sin^2\delta_k\left< \eta^{\dag}_{k-\pi} \eta_{k-\pi} \right> \right)\right.\cr
&+\left.(u^*_{k-\pi,k}u_{k-\pi,k-\pi}-u_{k-\pi,k}u^*_{k-\pi,k-\pi})\left(\sin^2\delta_k\left< \eta^{\dag}_k \eta_k \right>+\cos^2\delta_k\left< \eta^{\dag}_{k-\pi} \eta_{k-\pi} \right> \right)\right.\cr
&+\left.(u^*_{k,k}u_{k-\pi,k-\pi}-u_{k-\pi,k}u^*_{k,k-\pi})\frac{\sin 2\delta_k }{2}\left(\left< \eta^{\dag}_k \eta_k \right>-\left< \eta^{\dag}_{k-\pi} \eta_{k-\pi} \right> \right)\right.\cr
&+\left.(u^*_{k-\pi,k}u_{k,k-\pi}-u_{k,k}u^*_{k-\pi,k-\pi})\frac{\sin 2\delta_k }{2}\left(\left< \eta^{\dag}_k \eta_k \right>-\left< \eta^{\dag}_{k-\pi} \eta_{k-\pi} \right> \right)\right]\;.
\label{Sttd}
\end{split}
\end{align}
\end{widetext}

\begin{acknowledgments}
This work was supported by the National Research Fund under Grants No. K109577, No. K115959 and No. KKP-126749. H.R. extends thanks to the "Theoretical Physics Workshop" and F.I. and G.R. to the Saarland University for supporting their visits to Budapest and Saarbr\"ucken, respectively.
\end{acknowledgments}

\end{document}